\documentclass[10pt,journal,compsoc]{IEEEtran}
\usepackage{amsmath}
\usepackage{amssymb}
\usepackage{amsthm}
\usepackage{graphicx}
\usepackage{setspace}
\usepackage{mathtools}
\usepackage[ruled, vlined, linesnumbered]{algorithm2e}
\usepackage{color,soul}

\usepackage{cite}
\usepackage{subfigure}
\usepackage{stmaryrd}
\usepackage{makecell}
\usepackage{breqn}
\usepackage{stackengine}
\usepackage{verbatim} 
\usepackage{glossaries}
\usepackage{color,soul}
\usepackage{xcolor}

\usepackage{xspace}

\newcommand{\nonl}{\renewcommand{\nl}{\let\nl\oldnl}}
\makeatletter
\newcommand\approxsim{\mathchoice
  {\@approxsim {\displaystyle}      {1ex} }
  {\@approxsim {\textstyle}         {1ex} }
  {\@approxsim {\scriptstyle}       {.7ex}}
  {\@approxsim {\scriptscriptstyle} {.5ex}}}
\newcommand\@approxsim[2]{%
  \mathrel{%
    \ooalign{%
      $\m@th#1\sim$\cr
      \hidewidth$\m@th#1.$\hidewidth\cr
      \hidewidth\raise #2 \hbox{$\m@th#1.$}\hidewidth\cr
    }%
  }%
}
\usepackage{etoolbox}

\let\mybibitem\bibitem
\renewcommand{\bibitem}[1]{%
  \ifstrequal{#1}{nature}
    {\color{blue}\mybibitem{#1}}
    {\color{black}\mybibitem{#1}}%
}

\makeatother

\ifCLASSINFOpdf
 
\else
  
\fi

\begin{document}
  \title{ \huge Coordinated Half-Duplex/Full-Duplex Cooperative Rate-Splitting Multiple Access in Multi-Cell Networks}
  \author{Mohamed Elhattab, Shreya Khisa, Chadi Assi, Ali Ghrayeb, Marwa Qaraqe, Georges Kaddoum}
  \IEEEtitleabstractindextext{%
  \begin{abstract}
  This paper explores downlink Cooperative Rate-Splitting Multiple Access (C-RSMA) in a multi-cell wireless network with the assistance of Joint-Transmission Coordinated Multipoint (JT-CoMP). In this network, each cell consists of a base station (BS) equipped with multiple antennas, one or more cell-center users (CCU), and multiple cell-edge users (CEU) located at the edge of the cells. Through JT-CoMP, all the BSs collaborate to simultaneously transmit the data to all the users including the CCUs and {CEUs}. To enhance the signal quality for the CEUs, CCUs relay the common stream to the CEUs by operating in either half-duplex (HD) or full-duplex (FD) decode-and-forward (DF) relaying mode. In this setup, we aim to jointly optimize the beamforming vectors at the BS, the allocation of common stream rates, the transmit power at relaying users, i.e., CCUs, and the time slot fraction, aiming to maximize the minimum achievable data rate. However, the formulated optimization problem is non-convex and is challenging to solve directly. To address this challenge, we employ change-of-variables, first-order Taylor approximations, and a low-complexity algorithm based on Successive Convex Approximation (SCA). We demonstrate through simulation results the efficacy of the proposed scheme, in terms of average achievable data rate, and we compare its performance to that of four baseline schemes, including HD/FD cooperative non-orthogonal multiple access (C-NOMA), NOMA, and RSMA without user cooperation. {The results show that the proposed FD C-RSMA can achieve $25$\%  over FD C-NOMA and the proposed HD C-RSMA can achieve $19$\% over HD C-NOMA respectively, when the BS transmit power is $20$ dBm.} 
  \end{abstract}
\begin{IEEEkeywords}
Beamforming, Coordinated Multipoint, Cooperative RSMA, FD, HD, Multi-cell networks, User relaying, and 6G.   
\end{IEEEkeywords}}
 \maketitle
\section{Introduction}
\subsection{Background}
The sixth generation (6G) of cellular networks has been attracting substantial attention from industry and academia, surpassing the era of fifth generation (5G) networks \cite{9040264, 8869705}. 6G will have to cope with upsurging needs for ultra-high reliability, high throughput, diverse quality-of-service (QoS) requirements, ultra-low latency, and massive connectivity to fulfill the requirements of extremely ultra-reliable and low-latency communication (eURLLC), further-enhanced mobile broadband (FeMBB), and ultra massive machine type communication (umMTC) services \cite{9040264, 8869705}. In order to achieve these heterogeneous and strict requirements, enormous efforts have been recently devoted to improving the multiplexing gains of next-generation multiple access (NGMA) techniques  \cite{9831440, 10038476}. In consequence, rate-splitting multiple access (RSMA), which is built upon the rate-splitting (RS) principle, has recently emerged as a promising non-orthogonal transmission mechanism for multiple access and interference management in multi-antenna systems \cite{9831440, 10038476}. 

\par RSMA provides a more general and robust transmission framework in comparison with space division multiple access (SDMA) and non-orthogonal multiple access (NOMA). The main idea of RSMA is to partially treat the multi-user interference as noise and partially decode it. This can be achieved by splitting the messages of the user equipments (UEs) into private and common parts and then transmitting them using superposition coding (SC). In this transmission mechanism, the common message can be decoded by multiple UEs and then removed from the received signal by employing successive interference cancellation (SIC). Finally, each private message is only decoded by its intended UE. Thus, RSMA can overcome the limitations in existing schemes, such as SDMA, which fully treats multi-user interference as noise, NOMA, which forces all co-scheduled UEs to fully decode the multi-user interference belonging to other UEs using multi-level SIC at the receiver, and the inefficient orthogonal multiple access (OMA), which tackles the multi-user interference by allocating orthogonal resources among UEs.
\par It is worth mentioning that RSMA, specifically 1-layer RSMA, may suffer from a performance loss, which may limit its potential gain. This is due to the requirement for the decoding of the common stream by all users, which renders the achievable common rate to be constrained by the worst-case achieved rate of the common stream at all UEs. Consequently, in the case of heterogeneous channel strengths between the base station (BS) and the associated UEs, the gain that comes from the achievable common rate may vanish. To tackle this limitation and unleash the full potential of RSMA, the integration between cooperative relaying and RSMA has been introduced, which is known as cooperative RSMA (C-RSMA) \cite{9123680, 8846761, 9627180, 9771468}. Specifically, in C-RSMA, the system leverages the successive decoding property and the requirement for all UEs to decode the common stream. In particular, UEs with favorable channel conditions, known as cell-center users (CCUs), can assist the BSs by operating in either full-duplex (FD) or half-duplex (HD) decode-and-forward (DF) relaying mode to forward the decoded common message to the users with bad channel gains from the transmitting node, which is referred to as cell-edge users (CEUs). This will improve the signal quality at the CEUs, and hence, it helps in improving the achievable rate for decoding the common stream. It has been proven that C-RSMA can enlarge the rate region \cite{8846761}, improve user fairness \cite{9123680, 9771468}, minimize power consumption \cite{9627180}, extend network coverage \cite{9831440}, and enhance secrecy rate \cite{9217123} compared to the traditional non-cooperative 1-layer RSMA.

\subsection{Motivation and State-of-the-Art}
\par The potential of RSMA has been studied in many works, and has been compared to state-of-the-art multiple access schemes in single-cell wireless networks  \cite{7470942,9737523,8846706,9684855,9771468,10109654,9676684,9257190, 9217123} and in multi-cell wireless networks\cite{9445019,9257190,8756076,9573421,9759225,9896157,8756668,9195473}. 
 RSMA technology has been widely studied in single-cell networks in the context of multiple-input single-output (MISO) and multiple-input and multiple-output (MIMO) scenarios in terms of network spectral efficiency, energy efficiency, network coverage, and user fairness \cite{7470942,9737523,8846706,9684855,9771468,10109654,9676684,9257190, 9217123}. In \cite{7470942}, the authors focused on the challenges and opportunities associated with the use of RSMA in MIMO wireless networks. 
The authors in \cite{9737523} investigated RSMA-assisted cell-free massive MIMO in an imperfect channel state information (CSI) scenario for massive machine-type communications. Their results showed that RSMA is capable of mitigating pilot contamination while achieving higher gain than conventional cell-free massive MIMO system. The authors in \cite{8846706} studied RSMA for non-orthogonal unicast and multicast (NOUM) transmission in order to maximize both spectral and energy efficiencies. Using RS technique, they split unicast messages into common and private parts and encoded the common parts along with the multicast message into a super-common stream decoded by all users.  Numerical results showed that their proposed approach achieves higher gain in a wide range of user deployments with a diversity of channel directions, channel strengths, and channel qualities of CSI at the transmitter.  A sum-rate maximization problem was studied in \cite{9461768} RSMA for downlink which was one of the earliest attempts to show the performance gain of RSMA over other MA techniques. RSMA has also been applied to incorporate with the reconfigurable intelligent surface (RIS) to facilitate the signal quality of the users when there is a blockage between the transmitter and the receivers \cite{9508885,9854887}. However, these works are solely focused on general RSMA and do not involve user cooperation.
\par In order to improve the signal quality at the CEUs side, there are several works studied the potential of cooperative relaying through CCU to relay the common stream in both HD and FD mode \cite{9123680,9771468,10436906, 10109654, 10233866, 9627180}, which is called C-RSMA. Particularly, the authors in \cite{9123680} proposed a C-RSMA framework for $K$-users where they proposed a solution to find a strong user that can relay the common stream signal to the far users in HD mode. Meanwhile, the authors in \cite{9771468} studied FD C-RSMA in multi-group multicast scenarios. In \cite {10436906}, a two-user case for C-RSMA in an MISO scenario has been studied. This work has been extended in \cite{10109654,10233866} by integrating RIS with C-RSMA and simultaneous wireless information and power transfer (SWIPT). The authors in \cite{ 9627180} proposed a 
C-RSMA scheme in a
two-layer heterogeneous network (HetNet), aiming to maximize the downlink sum rate
of all small cell users. Note that,
to the best of our knowledge, the research mentioned above contributions of
the C-RSMA beamforming optimization problem in the context of single-cell
C-RSMA networks. However, the more challenging multi-cell
scenario was not addressed in literature in the context of C-RSMA.

\begin{table*}[htbp]
\caption{{Comparison table for C-RAN and JT-CoMP-based RSMA/C-RSMA-based works}}
\begin{center}
\footnotesize
\begin{tabular}{ |c|l| l|l|l|l|}
 \hline
  \textbf{Existing works} & \textbf{Summary}  & \textbf{JT-CoMP} & \textbf{RSMA} & {\makecell[l]{\textbf{User} \\ \textbf{cooperation}}} & \textbf{Multi-cell}\\
 
 \hline
 A. A. Ahmad et al.\cite{9445019} & {\makecell[l]{Studied a 1-layer RSMA-based \\ C-RAN framework}} & $\times$ & $\checkmark$ & $\times$ & $\checkmark$\\
 \hline
D.Yu et. al \cite{8756076}& {\makecell[l]{Proposed a hierarchical RSMA-assisted \\ C-RAN framework}} & $\times$ & $\checkmark$ & $\times$ & $\checkmark$\\
 \hline
 {\makecell[l]{K. Weinberger \\ et. al \cite{9573421}}} & {\makecell[l]{A RIS-assisted 1-layer RSMA-based \\ C-RAN framework was proposed}} & $\times$ & $\checkmark$ & $\times$ & $\times$\\
 \hline
 Q. Zhu et al. \cite{9896157}& {\makecell[l]{An 1-layer RSMA-enabled user-centric\\ RRH clustering in C-RAN was studied}} & $\times$ & $\checkmark$ & $\times$ & $\times$\\
 \hline
 Z. Zhou et al. \cite{9573421} & {\makecell[l]{RSMA for multigroup multicast beamforming \\in a 
cache-enabled C-RAN was studied}} & $\times$ & $\checkmark$ & $\times$ & $\times$\\
 \hline
 Y. Mao et.al \cite{8756668}& {\makecell[l]{Proposed a JT-CoMP based 1-layer RSMA framework}} & $\checkmark$ &$\checkmark$ & $\times$ & $\checkmark$ \\
 \hline
J. Zhang et.al \cite{9195473}& {\makecell[l]{Proposed a 1-layer RSMA-based JCT framework}}&
 $\times$ & $\checkmark$ & $\times$ & $\checkmark$\\
 \hline
 Our proposed work & {\makecell[l]{Proposed a 1-layer C-RSMA-based JT-CoMP framework}} & $\checkmark$ & $\checkmark$ & $\checkmark$ & $\checkmark$\\
 \hline
\end{tabular}
\end{center}
\label{table1}
\end{table*}
\par Most of the existing literature on RSMA$/$ C-RSMA has mainly focused on single-cell setups \cite{9123680,8846761,9771468,10436906, 9627180,  9461768, mao2018rate, 8352643,9382277}, where a single BS serves multiple users. However, in a multi-cell scenario, inter-cell interference (ICI) plays a crucial role. It can diminish the performance of CEUs, who coincide with far/weak users in the C-RSMA framework \cite{9123680,8846761,9771468,10436906, 9627180}. Specifically, the CEUs suffer from severe ICI, resulting in a low received signal-to-interference-plus-noise ratio (SINR), thereby limiting the benefits of RSMA/C-RSMA. Furthermore, the effect of ICI is exacerbated as the network becomes denser.
\par  A potential solution to the aforementioned challenge is to integrate RSMA/C-RSMA with Third Generation Partnership Project (3GPP) interference mitigation techniques, such as joint transmission coordinated multi-point (JT-CoMP) or enhanced ICI coordination (eICIC). Utilizing these techniques can improve the spectral efficiency of CEUs. Therefore, integrating JT-CoMP with C-RSMA-empowered cellular networks will create a promising framework for next-generation wireless communication. In particular, in C-RSMA-empowered wireless networks, CEUs can benefit from the coordination between BSs due to CoMP transmission and the cooperation between CCUs and the BS in each cell due to the C-RSMA technique. A few studies have recently focused on investigating RSMA in cloud radio access network (C-RAN) multi-cell networks. The authors in \cite{9445019} proposed an RSMA-based C-RAN framework to maximize the ergodic sum rate of the wireless network subject to per-BS transmit power and fronthaul capacity constraints in an imperfect CSI scenario. It is one of the earliest approaches to evaluate the performance of C-RAN with RSMA. The authors in \cite{8756076} proposed a  C-RAN framework utilizing the hierarchical RSMA technique where they used group common messages and private user messages to maximize the minimum rate of the network.  In \cite{9573421}, RSMA for multigroup multicast beamforming in a cache-enabled C-RAN was studied to maximize the minimum weighted rate among all users. To facilitate direct line-of-sight (LoS) communication between BS and the users, a RIS-assisted C-RAN framework with RSMA was proposed, considering an efficient user clustering methodology in \cite{9759225}. An analytical model for RSMA-enabled user-centric remote radio head (RRH) clustering in C-RAN was studied in \cite{9896157}. Meanwhile, a RSMA-based joint coordinated transmission (JCT) framework considering intra-cell and inter-cell interference management to maximize network energy efficiency was proposed in \cite{9195473}. The authors in \cite{8756668} explored the downlink RSMA JT-CoMP to maximize the weighted sum rate of the users. It is worth mentioning that all the aforementioned works \cite{9445019,9257190,8756076,9573421,9759225,9896157,8756668,9195473} on multi-cell networks have primarily focused on evaluating the effectiveness of integrating CoMP with RSMA without considering user cooperation. \textit{However, this integration may not perform optimally in scenarios where CEUs experience extremely weak channels or have high QoS requirements. Therefore, it is crucial to address and prioritize the enhancement of signal quality for CEUs \cite{10154612}.} To the best of our knowledge, no existing literature has explored the performance benefits of C-RSMA in a multi-cell scenario. These facts motivated us to study the performance of JT-CoMP with C-RSMA.
For the reader's convenience, we provided a comparison table in Table I in order to differentiate between the existing works and our proposed work.

\begin{table*}
\caption{{Summary of key symbols}}
\begin{center}
\begin{tabular}{ |c|l| }
 \hline
  \textbf{Symbol} & \textbf{Description}  \\
 \hline
 $N$, $N$, $K$, $L$, $M$ & {\makecell[l]{Number of BSs, number of cells,\\ number of CCUs, number of CEUs, the total number of users}}   \\
 \hline
$\mathcal{N}_b$,  {$\mathcal{N}_c$}, $\mathcal{K}$, $\mathcal{L}$, $\mathcal{M}$ & {\makecell[l]{Set of all BSs, set of all cells, \\ set of all CCUs, set of all CEUs, set of all users}}\\
 \hline
 $N_t$& Number of antennas at each BS\\
 \hline
 $\bar{\rm O}$ & Operating mode of the CCUs, i.e., HD, FD\\
 \hline
 $n_b, {n_c}, k, l, m$ & Index of the BS, cell, CCU, CEU, any user \\
 \hline
 $\textbf{p}_{n_b,c}$, $\textbf{p}_{n_b,m}$ & {\makecell[l]{Beamforming/precoding vectors of BS-$n_b$ to transmit common stream $s_c$ \\ and private stream $s_m$ of user-$m$}} \\
 \hline
 $\textbf{w}_{c}$, $\textbf{w}_{m,p}$ & {\makecell[l]{JT-CoMP beamforming/precoding vectors to transmit common stream $s_c$ \\ and private stream $s_m$ of user-$m$}} \\
 \hline
 $P_{k}^{\rm {tot}}$ & Power budget of CCU-$k$\\
 \hline
 $\textbf{v}^{\bar{\rm O}}$ & Distributed beamforming vector\\
 \hline
 $v_{k}$ & Weighing element of CCU-$k$\\
 \hline
\makecell{$\textbf{h}_{n_b,m}$,\\
$\textbf{h}_{n_b}$,\\
$\textbf{h}_{l}$,\\$\textbf{h}_{l,d}$,\\
$h_{k,l}$, \\
$h_{k,\rm SI}$} & {\makecell[l]{Channel coefficient  of BS-$n_b$ $\rightarrow$ user-$m$, \\
channel vector between all BSs $\rightarrow$ CCU-$k$, \\channel vector between all BSs  $\rightarrow$ CEU-$l$, 
\\channel vector between all CCUs $\rightarrow$ CEU-$l$,\\
channel coefficient between CCU-$k$ $\rightarrow$ CEU-$l$, \\SI channel coefficient of CCU-$k$}}\\
 \hline
 $\theta$ & Time slot duration \\
 \hline
 $F_{n_b}^{\rm Fh}$ & Fronthaul capacity of BS-$n_b$  \\
 \hline
 $P_{n_b}^{\rm \max}$ & Power budget of BS-$n_b$\\
 \hline
\end{tabular}
\end{center}
\label{table1}
\end{table*}
\subsection{Contributions}
{Our main focus of this work is to evaluate and characterize the potential gains of the CoMP-assisted
C-RSMA framework in a multi-cell scenario in comparison with
other multiple access techniques developed in the literature.} Hence,
this paper addresses the resource management problem in CoMP-assisted FD/HD C-RSMA-enabled cellular networks, which has not been investigated before. Another novel aspect of the paper is that it is one of the early attempts to explore the performance of C-RSMA in multi-cell wireless networks. To this end, we summarize the contributions of the paper as follows.


\begin{itemize}
\item 
{
A downlink JT-CoMP-assisted HD/FD C-RSMA in a multi-cell framework, comprising a BS equipped with multiple antennas, multiple CCUs, and multiple CEUs positioned at the intersection area of different cells, is considered. To enhance CEUs' signal quality, CCUs collaboratively construct a distributed beamformer and act as an HD or FD relay, forwarding the decoded common stream to assist the CEUs. Our proposed 1-layer C-RSMA remains unaffected by inter-user interference, as only a common stream is relayed. In contrast, in the C-NOMA multi-user scenario, each CCU relays the signal of its paired CEU, leading to the relay of different signals during the cooperative phase, causing inter-user interference. }
\item We formulate an optimization problem to jointly design the private and common beamforming vectors along with the common stream rate allocation, the transmit power at the relaying users, and the time slot duration to maximize the minimum achievable data rate in the multi-cell JT-CoMP framework. In this framework, we consider the power budget at BSs, transmission power constraints at user relaying, and fronthaul capacity constraints at the BSs. 
\item Our formulated problem is a non-convex quadratically constrained quadratic programming (QCQP) problem whose complexity is non-deterministic polynomial-time (NP)-hard. To tackle this challenge, we first transform the original problem into a convex one by invoking several auxiliary variables and first-order Taylor approximations. Then, we propose a successive convex approximation (SCA)-based, low-complexity, efficient iterative algorithm to solve the transformed problem efficiently.
\begin{figure}[!t]
    \centering
\includegraphics[scale=0.2]{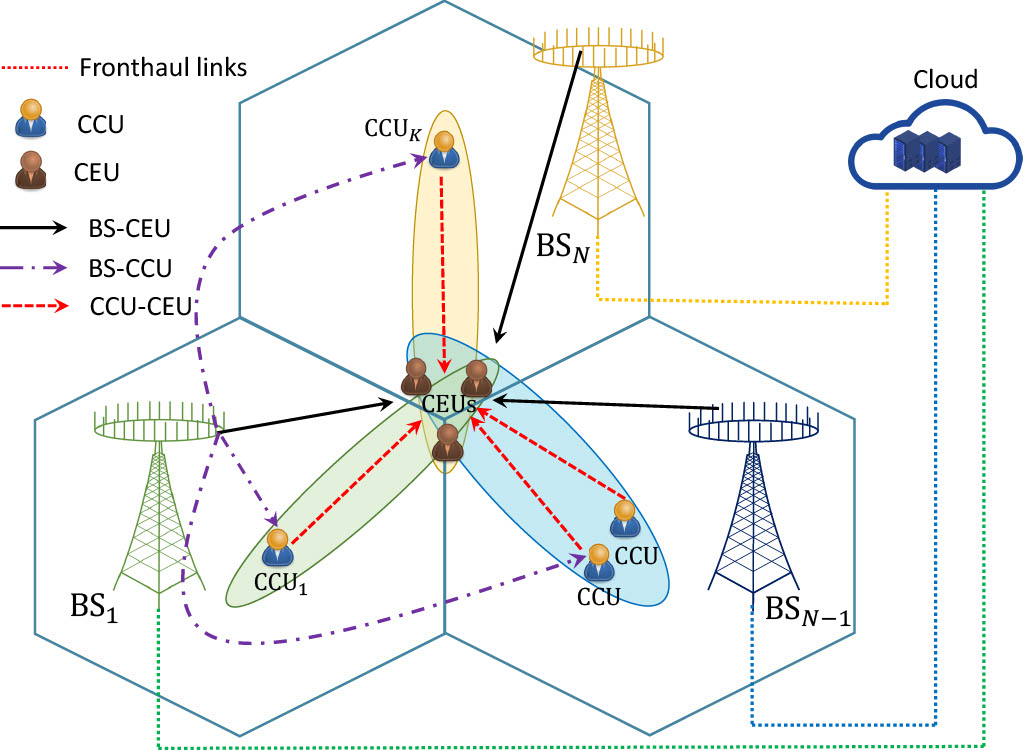}
    {\caption{{Proposed CoMP-assisted C-RSMA system model.}}}
    \label{System_Model_1}
\end{figure}
\item We conducted extensive simulations to assess the performance of the proposed JT-CoMP-enabled HD/FD C-RSMA multicell networks. To validate the effectiveness of our framework, we compare its performance against four baseline schemes. Additionally, we evaluate the performance of these schemes under varying system parameters, including self-interference (SI) channel gain, varying power levels for both BSs and UEs, diverse channel disparities, and various ranges of fronthaul capacities.  
\end{itemize}

\subsection{Notations and Paper Organization}
\par In this paper, vectors, and matrices are denoted by bold-face letters.  For a complex-valued vector $\boldsymbol {y}, |\boldsymbol {y}|$ accounts for its Euclidean norm. 
For any general matrix $\boldsymbol {M}, \boldsymbol {M}^{H}$ and $\boldsymbol {M}^{T}$ denote its conjugate transpose and its transpose, respectively. Finally, for a complex number $z$, $z^{\dag}$ and $\tt{Re}\{z\}$ represent its complex conjugate and its real part, respectively. 
\par {The notations adopted throughout the paper are presented in Table  2.} The remainder of this paper is organized as follows. Section 2 introduces the system model. Section 3 presents the achievable rate analysis for both HD and FD relaying modes. Sections 4 and 5 describe the formulated problems. and their solutions. Section 6 presents the simulation results, and finally, we conclude in Section 7.

\section{System Model}
\subsection{Network Model}
We consider a downlink transmission in a multi-user multi-cell network consisting of $N$ cells, $N$ BSs, $K$ CCUs, and $L$ CEUs, where $L \le K$ and $K \ge N$ as depicted in Fig. 1. We assume that $K$ CCUs are distributed over the $N$ cells. For notation convenience, $\mathcal{N}_c \triangleq \{1, 2, \dots, N\}, \mathcal{N}_b \triangleq \{1, 2, \dots, N\}, \mathcal{K} \triangleq \{1, 2, \dots, K\}$, $\mathcal{L} \triangleq \{1, 2, \dots, L\}$, and $\mathcal{M} \triangleq \{1, 2, \dots, M\}$ define the set of cells, the set of BSs, the set of the CCUs, the set of CEUs, and the set of UEs, respectively. Moreover, $N$, $N$, $K$, $L$, and $M$ denote the cardinality of $\mathcal{N}_c, \mathcal{N}_b$, $\mathcal{K}$, $\mathcal{L}$, and $\mathcal{M}$, respectively. We assume that each cell is equipped with one BS and one or more than one CCU. In addition, we assume that each BS has $N_t > 1$ transmit antennas, and each user is equipped with a single antenna. {In our proposed system model, we assume an underloaded RSMA-enabled wireless network scenario where the number of antennas at the BSs is greater than or equal to the number of UEs served \cite{mao2018rate}.}  Without loss of generality, we assume that CCU-$k\in \mathcal{K}$, which belongs to cell-$n_c \in \mathcal{N}_c$, has a strong channel gain with BS-$n_b$. Meanwhile, the CEUs experience less distinctive channel gains from the BSs, i.e., suffer from severe propagation loss due to their distant locations from the adjacent BSs. Finally, to avoid the ICI due to the co-channel deployment, we focus on a fully cooperative multi-cell network in which all the BSs serve all the $M$ users through JT CoMP \cite{7839266, 8756668}. Moreover, all the wireless channels are assumed to be independent and follow a Rayleigh distribution. \footnote{{We assume that full CSI is available for all wireless links at the
centralized pool. This setup allows the optimization to be performed centrally, leveraging the
powerful computational resources at the baseband units (BBU) pool, ensuring efficient and
low-latency solutions.}} Hence, the channel gains for those wireless links follow Exponential distributions with parameters $\lambda_s$ (BS-$n_b\rightarrow$ CCU-$k)$, $\lambda_w$ (BS-$n_b\rightarrow$ CEU-$l)$ and $\lambda_d$ (CCU-$k\rightarrow$ CEU-$l)$, respectively.
\par In this model, all BSs are connected to central processing (CP) through capacity-limited fronthaul links, denoted as $F_{n_b}^{\rm{FH},b}$, for information and data exchange among different BSs so that each user, i.e., CCU or CEU, can be simultaneously served by all the BSs \cite{8756668}. In the considered system model, we apply the RS approach, where each user's message is divided into two parts, i.e., private and common \cite{9831440}. Specifically, the private message for each user is encoded into a private stream, while all the common parts of all users are encoded together into a common stream using a codebook that is shared by all users \cite{9831440}.  Furthermore, to improve the system performance, all CCUs act as FD or HD DF relays to assist the BSs in their transmissions to the CEUs \cite{8846761, 9123680, 9627180}. Specifically, the transmission model consists of two main phases, i.e., the direct transmission (DT) phase and the cooperative transmission (CT) phase \cite{8846761, 9123680, 9627180}. In the DT phase, BSs transmit signals to all users, meanwhile, in the CT phase, all the CCUs cooperatively constitute a distributed beamformer of the decoded common stream to assist the CEUs. The detailed transmission model is discussed in the following section.
\subsection{Transmission Model}
The DT and CT phases for the proposed multi-user multi-cell are detailed as follow.
\par \textit{1) DT Phase:} In this phase, all BSs serve the $M$ users. Following the principle of $1$-layer RSMA, the message $A_m$ intended for user$-m$, $\forall m \in \mathcal{M}$ is split into a private part $A_{m,p}$ and a common part $A_{m,c}$ \cite{9831440}. The common parts of all users. {$A_{1,c}, \dots, A_{K,c}, A_{1,c}, \dots, A_{L,c}$} are combined into a common message $A_c$ \cite{9831440}. It is then encoded into a common stream $s_c$ using a codebook that is shared by all users. Note that this $s_c$ should be decoded by all users since it includes parts of the messages of all users, i.e., CCUs and CEUs. Meanwhile, the private part $A_{m,p}$, $\forall m \in \mathcal{M}$, of user-$m$ is independently encoded into private stream denoted as $s_m$, $\forall m \in \mathcal{M}$, which is only decoded by user-$m$. It is noteworthy that all streams $s_c$ and $\{s_m| m\in \mathcal{M}\}$ have normalized power, i.e., $\mathbb{E}[|s_c|^2] = \mathbb{E}[|s_m|^2] = 1$. The encoded streams are linearly precoded by each BS and then broadcasted to the users. The encoded stream at BS-$n_b$ is given by 
\begin{equation}
    {\label{BF}
    \textbf{x}_{n_b}^{[1]} = \textbf{P}_{n_b}^{\rm{\Bar{O}}}\textbf{s} = \textbf{p}_{n_b,c}^{\rm{\Bar{O}}} s_c + \sum_{m \in \mathcal{M}} \textbf{p}_{n_b,m}^{\rm{\Bar{O}}} s_m, \qquad \forall n_b \in \mathcal{N}_b,}
\end{equation}
where ${\rm{\Bar{O}}} \in \{\rm{HD, FD}\}$ defines the operating mode of the CCUs, $\textbf{s}=\left[s_c, s_1, \dots, s_{M}\right]^T$, $\textbf{P}_{n_b}^{\rm{\Bar{O}}} = \left[\left(\textbf{p}_{n_b,c}^{\rm{\Bar{O}}}\right)^T, \left(\textbf{p}_{n_b,1}^{\rm{\Bar{O}}}\right)^T, \dots,\left(\textbf{p}_{n_b,M}^{\rm{\Bar{O}}}\right)^T\right],$ is the precoding matrix and $\textbf{p}_{n_b,c}^{\rm{\Bar{O}}}, \textbf{p}_{n_b,m}^{\rm{\Bar{O}}} \in \mathbb{C}^{1 \times N_t }, \forall n_b \in \mathcal{N}_b, m \in \mathcal{M}.$ Note that the precoding vectors inside the precoding matrix are required to be orthogonal/semi-orthogonal to each other to ensure that the transmitted signals do not interfere with each other. Given normalized symbol power, the transmit power at BS-$n_b$ can be given as $\mathrm{tr}\left(\textbf{P}_{n_b}^{\rm{\Bar{O}}}(\textbf{P}_{n_b}^{\rm{\Bar{O}}})^H\right)$. Since each BS has a limited power budget $P_{n_b}^{\rm \max}$, we have the following power budget constraint $\mathrm{tr}\left(\textbf{P}_{n_b}^{\rm{\Bar{O}}}(\textbf{P}_{n_b}^{\rm{\Bar{O}}})^H\right) \leq P_{n_b}^{\rm \max}$. In addition, the CCUs receive the superimposed signals transmitted by the BSs and initiate the process of decoding the common stream $s_c$ by considering the interference from the private streams as noise. Then, through SIC, the decoded common stream is removed from the received signal. Afterward, each CCU decodes its private stream, considering the remaining interference from the other private messages as noise.
\par \textit{2) CT Phase:} In this phase, all the CCUs cooperate to assist in transmitting the common stream to the CEUs. Specifically, all the CCUs cooperate and act as a virtual antenna array to construct a distributed beamformer to improve the performance of the CEUs, and, consequently, the performance of the network in terms of system spectral efficiency. Based on the above discussion, the transmit signal from all CCUs, i.e., the virtual distributed antenna array, in the CT phase can be written as 
\begin{equation}
    \textbf{x}^{[2]} = \textbf{v}^{\rm{\Bar{O}}}s_c,
\end{equation}
where $\textbf{v}^{\rm{\Bar{O}}} = \left[v_1^{\rm{\Bar{O}}}, \dots, v_K^{\rm{\Bar{O}}}\right]^T$ is the distributed beamforming vector, and $v_{k}^{\rm{\Bar{O}}}$ is the weighting element of CCU-$k$. On the CEUs side, the common stream can be received from two different sources: the first source is the CCUs transmission, and the second source is the transmission of the BSs. Finally, the CEUs combine the different copies of the common stream to start the decoding process. {Since the common stream encapsulates information intended for all users and undergoes decoding by each user, and the performance of the weakest user significantly influences the achievable common stream rate it is both practical and advantageous to propagate the common stream to the CEUs. It is noteworthy that it is a widely adopted technique in cooperative communication in the context of RSMA \cite{9123680, 8846761, 9627180, 9771468}.}

Note that the advantage of 1-layer C-RSMA stems from its immunity to inter-user interference
during the CT phase, as the same signal is relayed to all CEUs. In addition, the joint relaying
towards the CEUs from all the CCUs results in improved signal quality at the CEUs end. Moreover, the CCUs can operate in either HD or FD relaying mode. Regarding the HD relaying mode, the DT and CT phases occur in two consecutive time slots \cite{9123680}. Meanwhile, for the FD scenario, they occur in the same time slot at the cost of residual SI at the CCUs \cite{9627180, 9771468}. After presenting the main operations of multi-user multi-cell C-RSMA, we derive the received SINR and the corresponding achievable data rates at the CCUs and CEUs for the two considered relaying modes in the next section.
\section{Achievable Data Rate Analysis}
\subsection{HD Relaying Mode}
In the HD relaying mode, in the first time slot, the BSs transmit the superimposed signals to all users. In the second time slot, the BSs are silent, and the CCUs transmit the common stream to the CEUs. Let $\theta \in [0,1]$ be the fraction of time that is assigned to the DT phase; while the rest $(1- \theta)$ is allocated to the CT phase. Consequently, the received signal at user-$m$, in the first time slot can be expressed as
\begin{equation}
    y_m^{[1]} = \sum_{n_b\in\mathcal{N}_b}\textbf{h}_{n_b,m}^H \textbf{x}_{n_b}^{[1]} + z_m, \qquad \forall m \in \mathcal{M},
\end{equation}
where $z_m$ is the additive white Gaussian noise (AWGN) with zero mean and variance $\sigma^2$, i.e., $z_m \sim \mathcal{CN}(0, \sigma^2)$. As discussed earlier, the CCUs start by decoding the common stream, and hence, the achievable data rate at CCU-$k$, $\forall k \in \mathcal{K}$, to decode the common stream in the first time slot is given by
\begin{align}\footnotesize
  \mathcal{R}^{[1], \rm{HD}}_{k,c} &= \theta \log_2\left(1 +  \frac{\textbf{h}_{k}\textbf{w}_c\textbf{w}_c^H\textbf{h}_{k}^H}{ \textbf{h}_{k}\left(\sum_{m\in\mathcal{M}}\textbf{w}_{m,p}\textbf{w}_{m,p}^H\right)\textbf{h}_{k}^H + \sigma^2}\right),
\end{align}
where $\textbf{h}_{k} = \left[\textbf{h}_{1,k}^H,\dots,\textbf{h}_{K,k}^H\right], \textbf{w}_c = \left[\textbf{p}_{1,c}^{\rm{HD}},\dots, \textbf{p}_{K,c}^{\rm{HD}}\right]^T$, and $\textbf{w}_{m,p} = \left[\textbf{p}_{1,m}^{\rm{HD}}, \dots, \textbf{p}_{K,m}^{\rm{HD}}\right]^T$. After removing the common stream from the received signal, the achievable data rate at CCU-$k, \forall k \in \mathcal{K}$ for the private stream can be expressed as 
\begin{align}\footnotesize
   \mathcal{R}_{k, p}^{[1], \rm{HD}} &= \theta \log_2\left(1 + \right. \notag \\ & \left.  \frac{\textbf{h}_{k}\textbf{w}_{k,p}\textbf{w}_{k,p}^H\textbf{h}_{k}^H}{\textbf{h}_{k}\left(\sum_{j\in\mathcal{M}, j \ne k}\textbf{w}_{j,p}\textbf{w}_{j,p}^H\right)\textbf{h}_{k}^H + \sigma^2}\right).
\end{align}
Note that the CEUs do not start the decoding process until they receive the common stream from the CCUs in the CT phase. Specifically, in the HD relaying mode, we assume that the CCUs employ a non-regenerative decode-and-forward (NDF) protocol to transmit the common stream $s_c$ to the CEUs \cite{9123680}. In particular, each CCU re-encodes $s_c$ with a codebook that is independently generated from that of the BSs \cite{9123680}. Then, the CCUs cooperatively forward the common stream $s_c$ using a distributed beamforming vector $\textbf{v}^{\rm{HD}}$ in which each CCU contributes with a factor $v_{k}^{\rm{HD}}$. Here, the transmit power of each CCU-$k$ is constrained by its power budget denoted as $P_{k}^{\rm{tot}}$, i.e., $|v_{k}^{\rm{HD}}|^2\leq P_{k}^{\rm{tot}}$. Consequently, the received signal at the CEU-$l$ in the CT phase is given by 
\begin{equation}
    y_{l}^{[2]} = \textbf{h}_{{l,d}}^H \textbf{x}^{[2]},
\end{equation}
where $\textbf{h}_{l,d} = \left[h_{1,1},\dots, h_{K, L}\right]^T$, where $h_{k, l}$ is the channel gain between CCU-$k \in \mathcal{K}$ and CEU-$l \in \mathcal{L}$.  Consequently, the data rates of decoding the common stream $s_c$ at the CEU-$l$, $\forall l \in \mathcal{L}$ in the first and second-time slots can be, respectively, given by
\begin{align}
    \mathcal{R}^{[1], \rm{HD}}_{l,c} & = \theta \log_2\left(1 + \frac{\textbf{h}_{l}\textbf{w}_c\textbf{w}_c^H\textbf{h}_{l}^H}{ \textbf{h}_{l}\left(\sum_{m\in\mathcal{M}}\textbf{w}_{m,p}\textbf{w}_{m,p}^H\right)\textbf{h}_{l}^H + \sigma^2}\right), \notag \\
    \mathcal{R}^{[2], \rm{HD}}_{l,c} &= (1 - \theta) \log_2\left(1 + \frac{|\textbf{h}_{l,d}^H \textbf{v}^{\rm{HD}}|^2}{\sigma^2}\right),
\end{align}
where $\textbf{h}_l = \left[\textbf{h}_{1,l}^H,\dots,\textbf{h}_{K,l}^H\right]$.
Note that the CEUs combine the decoded common stream in the DT and CT phases, and hence, according to \cite{9123680}, the achievable data rate for decoding the common stream at the CEU-$l$ can be expressed as 
\begin{equation}
    \mathcal{R}_{l,c}^{\rm{HD}} = \mathcal{R}^{[1], \rm{HD}}_{l,c} + \mathcal{R}^{[2], \rm{HD}}_{l,c}, \quad \forall l \in \mathcal{L}. 
\end{equation}
On the other hand, the achievable data rate at the CEU-$l$ to decode its private stream can be expressed as
\begin{align}
    \mathcal{R}_{l, p}^{[1], \rm{HD}} & = \theta \log_2\left(1 +  \frac{\textbf{h}_{l}\textbf{w}_{l,p}\textbf{w}_{l,p}^H\textbf{h}_{l}^H}{\textbf{h}_{l}\left(\sum_{j\in\mathcal{M}, j \ne l}\textbf{w}_{j,p}\textbf{w}_{j,p}^H\right)\textbf{h}_{l}^H + \sigma^2}\right).
\end{align}

 Common parts from all UEs are encoded together and generate a single common stream. Hence, the common stream contains information from all UEs in the system. Therefore, all UEs need to decode the common stream. To ensure successful decoding of the common stream by
all UEs, i.e., CCUs and CEUs, the achievable data rate
of the common streams should be the minimum among
the achievable common rates of all UEs, which can be
expressed as
 \begin{align}\label{eqn:12}
\mathcal{R}_{c}^{\rm{HD}} &= \min_m \left\{\mathcal{R}_{m,c}^{\rm {HD}}| m \in \mathcal{M}\right\}.
\end{align}

Since the common stream is formed by taking a portion of each UE's message, the achievable rate for the common stream should be shared among all the UEs. Thus, we define $C_m^{HD}$ as the portion of the common stream rate $R_c^{HD}$ that is assigned to UE-$m$, $\forall m \in \mathcal{M}$ and the sum of all portions of common parts from all UEs constitute the rate of the common stream, which is represented by, $\sum_m^M{C_m^{HD}}=R_c^{HD}$. Thus, we can obtain the total achievable rate for UE-$m$ as
\begin{equation}\label{eqn:13}
\mathcal{R}_m^{\rm{HD}} =  C_m^{\rm{HD}} + \mathcal{R}_{m, p}^{[1], \rm{HD}}, \qquad \forall m \in\mathcal{M}.
\end{equation}
\vspace{-0.65cm}
\subsection{FD Relaying Mode}
For the FD-coordinated C-RSMA, both the DT and CT phases are simultaneously executed using the same resource. For that reason, each CCU-$k$ suffers from an SI resulting from transmitting data to the CEUs and receiving data from the BSs simultaneously within the same channel use. Hence, the received signal at CCU-$k \in \mathcal{K}$, is given by
\begin{align}
    y_{k} & = \sum_{n_b \in \mathcal{N}_b} \textbf{h}_{n_b,k}^H\left(\textbf{p}_{n_b,c}^{\rm{FD}} s_c + \sum_{m \in \mathcal{M}} \textbf{p}_{n_b,m}^{\rm{FD}} s_m\right) \notag \\ & + h_{k,\rm{SI}}^H v_{k}^{\rm{FD}}  \tilde{s}_c + z_{k},
\end{align}
where $h_{k,\rm SI}$ is assumed to follow a complex symmetric Gaussian random variable with zero mean and variance $\Omega^2_{\rm{SI}}$, i.e., $\mathcal{CN}(0, \Omega^2_{\rm{SI}})$. Note that due to the processing time for the decoding process of the common stream $s_c$ at CCU-$k$, the message $s_c$ relayed from each CCU-$k$ to the CEUs is a delayed version of the actual common stream $s_c$ intended for the CEUs. In other words, by denoting $\tau$ as the common stream at the CCUs, the common stream $\Tilde{s}_c$ satisfies $\Tilde{s}_c(i) = s_c(i-\tau)$ decoding time, where $i$ denotes the $i$th time slot \cite{8302918, 8094966}. Based on the above discussion, the achievable data rate at CCU-$k$, $\forall k \in \mathcal{K}$, to decode the common stream is given by
\begin{align}   \mathcal{R}^{\rm{FD}}_{k,c} & = \log_2\left(1 + \right. \notag \\ & \left.  \frac{\textbf{h}_{k}\bar{\textbf{w}}_c\Bar{\textbf{w}}_c^H\textbf{h}_{k}^H}{ \textbf{h}_{k}\left(\sum_{m\in\mathcal{M}}\Bar{\textbf{w}}_{m,p}\bar{\textbf{w}}_{m,p}^H\right)\textbf{h}_{k}^H + |h_{k,\rm{SI}}^H v_{k}^{\rm{FD}}|^2 + \sigma^2}\right),
\end{align}
where $\bar{\textbf{w}}_c = \left[\textbf{p}_{1,c}^{\rm{FD}},\dots, \textbf{p}_{K,c}^{\rm{FD}}\right]^T$, and $\bar{\textbf{w}}_{m,p} = \left[\textbf{p}_{1,m}^{\rm{FD}}, \dots, \textbf{p}_{K,m}^{\rm{FD}}\right]$. Once $s_c$ is successfully decoded at CCU-$k$ and removed from the received signal $y_{k}$, the intended private stream $s_{k}$ can be decoded by treating all other private streams intended for other UEs as noise. Thus, the achievable data rate at CCU-$k$,$\forall k \in \mathcal{K}$, to decode its private stream can be expressed as
\begin{align}
    \mathcal{R}_{k, p}^{\rm{FD}} & = \log_2\left(1 + \right. \notag \\ & \left. \frac{\textbf{h}_{k}\bar{\textbf{w}}_{k,p}\bar{\textbf{w}}_{k,p}^H\textbf{h}_{k}^H}{\textbf{h}_{k}\left(\sum_{j\in\mathcal{M}, j \ne k}\bar{\textbf{w}}_{j,p}\bar{\textbf{w}}_{j,p}^H\right)\textbf{h}_{k}^H + |h_{k,\rm{SI}}^H v_{k}^{\rm{FD}}|^2 + \sigma^2}\right),
\end{align}
Afterward, all the CCUs forward the common stream $s_c$ to all the CEUs. Therefore, the received signal at the CEU-$l$, $\forall l \in \mathcal{L}$, can be given by
{\begin{align}
    y_{l} = \sum_{n_b \in \mathcal{N}_b} \textbf{h}_{n_b,l}^H\left(\textbf{p}_{n_b,c}^{\rm{FD}} s_c + \sum_{m \in \mathcal{M}} \textbf{p}_{n_b,m}^{\rm{FD}} s_m\right) \notag\\+ \sum_{k \in \mathcal{K}} h_{k, l}^{\dag} v_{k}^{\rm{FD}} \tilde{s}_c + z_l, \label{Eq: Rx CEU}
\end{align}}where the first term in \eqref{Eq: Rx CEU} stems from the transmission of the BSs, whereas the second term results from the cooperative transmissions of the CCUs. Note that the processing delay $\tau$ is very small in comparison with the time slot duration, i.e., $\tau < t_{i + 1} - t_i$, where $t_{i + 1}$ and $t_i$ are the time of the $(i+1)$th and $i$th time slots, respectively \cite{8302918,8094966}. Thus, the CEUs receive the common stream from the BSs and from the CCUs at approximately the same channel. Based on \cite{8302918, 8094966}, these two signals can be fully resolvable at the CEUs, and thus, they can be appropriately aligned and co-phased, and then combined using the maximum ratio combing (MRC) scheme. Based on that and on the results of \cite{8302918, 8094966}, the achievable data rate of the CEU-$l$, $\forall l \in \mathcal{L}$, to decode the common stream can be expressed as
{\begin{align}
    & \mathcal{R}^{\rm{FD}}_{l,c}  = \log_2\left(1 + \notag \right. \\ & \left. \frac{\textbf{h}_{l}\bar{\textbf{w}}_c\bar{\textbf{w}}_c^H\textbf{h}_{l}^H}{ \textbf{h}_{l}\left(\sum_{m\in\mathcal{M}}\bar{\textbf{w}}_{m,p}\bar{\textbf{w}}_{m,p}^H\right)\textbf{h}_{l}^H + \sigma^2} + \frac{|\textbf{h}_{l,d}^H \textbf{v}^{\rm{FD}}|^2}{\sigma^2}\right). 
\end{align}}
Meanwhile, the achievable rate at the CEU-$l$, $\forall l \in \mathcal{L}$,  to decode its private stream after successfully decoding the common stream can be expressed as
{\begin{equation}
    \mathcal{R}_{l, p}^{\rm{FD}} = \log_2\left(1 +\frac{\textbf{h}_{l}\bar{\textbf{w}}_{l,p}\bar{\textbf{w}}_{l,p}^H\textbf{h}_{l}^H}{\textbf{h}_{l}\left(\sum_{j\in\mathcal{M}, j \neq l}\bar{\textbf{w}}_{j,p}\bar{\textbf{w}}_{j,p}^H\right)\textbf{h}_l^H + \sigma^2}\right).
\end{equation}}As explained in the HD relaying mode scenario, the achievable data rate of the common streams should be the minimum between the achievable common rates of all users, which can be expressed as
\begin{align}
\mathcal{R}_{c}^{\rm{FD}} = \min_m \left\{\mathcal{R}_{m,c}^{\rm {FD}}| m \in \mathcal{M}\right\},
\end{align}
and the total achievable rate for user-$m$ can be obtained as
\begin{equation}
\mathcal{R}_m^{\rm{FD}} =  C_m^{\rm{FD}} + \mathcal{R}_{m, p}^{\rm{FD}}, \qquad \forall m \in\mathcal{M},
\end{equation}
 where $C_m^{\rm{FD}}$ represents the fraction of $\mathcal{R}_c^{\rm{FD}}$ assigned to user-$m$, $\forall m \in \mathcal{M}$ in the FD relaying mode scenario.

\section{Coordinated HD C-RSMA: Problem Formulation and Solution Approach}
\subsection{Problem Formulation}
For the sake of improving the network performance in terms of user fairness, we investigate the joint optimization of beamforming vectors at the BSs $\left(\textbf{P}^{\rm{HD}}\right), \forall n_b\in \mathcal{N}_b$, the distributed beamforming at the CCUs $\left(\textbf{v}^{\rm{HD}}\right)$, the common stream split $\left(\textbf{c}^{\rm{HD}}= \{C_1^{\rm{HD}}, \dots, C_{M}^{\rm{HD}}\}\right) $, and the time slot allocation $(\theta)$.\footnote{Once the optimization problem is solved using the available CSI, the centralized pool forwards the beamforming weights to the BSs through the fronthaul links. These BSs, in turn, forward the beamforming weights to the CCUs.} This framework is formulated as an optimization problem aiming to maximize the minimum user data rate. Based on the above discussion, the max-min rate optimization problem for the proposed coordinated C-RSMA system can be formulated as  
\allowdisplaybreaks
\begingroup
\begin{subequations}
\label{prob: Main1}
\begin{align}
\mathcal{P}-\mathrm{HD}: &\quad \max_{\substack{\textbf{P}^{\rm{HD}}, \label{P1}
\textbf{v}^{\rm{HD}},\\\textbf{c}^{\rm{HD}},\theta}} \min \mathcal{R}_m^{\rm{HD}} \quad \,\,\, m \in \mathcal{M}, \\
\text{s.t.}\,\, & \quad \,\,\, \sum_{m\in\mathcal{M}} C_m^{\rm{HD}} \leq R_c^{\rm{HD}}, \label{P1_C1}\\
&\quad \,\,\, \sum_{m\in\mathcal{M}} \mathcal{R}_m^{\rm{HD}} \leq F^{\rm{Fh}}_{n_b} && n_b \in \mathcal{N}_b, \label{P1_C2}\\
&\quad \,\,\, \mathrm{tr}\left(\textbf{P}_{n_b}^{\rm{HD}}(\textbf{P}_{n_b}^{\rm{HD}})^H\right) \leq P_{n_b}^{\rm \max}, && n_b \in \mathcal{N}_b,\label{P1_C3}\\
&\quad \,\,\, C_m^{\rm{HD}} \geq 0, && m \in \mathcal{M}, \label{P1_C4} \\
&\quad \,\,\, |v_{k}^{\rm{HD}}|^2\leq P_{k}^{\rm{tot}}, && k \in \mathcal{K},\label{P1_C5}\\
&\quad \,\,\, 0\leq \theta\ \leq 1, \label{P1_C6}
\end{align}
\end{subequations}
\endgroup
 where constraint \eqref{P1_C1} ensures the successful decoding of the common stream by all users, constraint \eqref{P1_C2} ensures that the achievable transmission rate for all users is limited by the finite fronthaul capacity, constraint \eqref{P1_C3} refers to the transmit power constraints at the BSs, and constraint \eqref{P1_C4} is the power budget at the CCUs.
It is important to mention that, and different from the literature, the formulated optimization problem at hand differs from traditional single-cell C-RSMA/RSMA rate problems in the C-RSMA system model in three ways: 1) the distributed beamforming vector applied by the multiple CCUs to assist in the transmission of the common stream to the CEUs, 2) the finite fronthaul capacity that is imposed due to the fully coordinated model, and 3) the per-BS transmit power constraint, which impacts the fraction of power allocated to each user from each BS. Due to the co-existence and high coupling of the optimization decision parameters, one can see that both the objective function and the constraints of the proposed optimization problem are non-convex. In other words, it can be easily seen that this optimization problem is a non-convex optimization problem, which is difficult to solve. 
\par One vital component in the optimization problem in Eqn. \eqref{P1} is the availability of the CSI for all considered wireless links to execute the optimization in a centralized manner. In the proposed model, we should estimate the CSI between the BSs and the users and the CSI between the CCUs and the CEUs. First, it is noteworthy that the CSI between the BSs and the users can be measured using one of the CSI acquisition techniques for multi-user wireless communication systems. For instance, the CSI reference signal (CSI-RS) in 5G new-radio (5G-NR), or for a more rigorous CSI estimation, the uplink sounding reference signal transmitted by all users is used to estimate the required channel gains for
the wireless communication links between the BSs and the cellular users in real-time as in LTE-A standard or as in 5G NR \cite{access2012radio}. Nevertheless, considering the SI and device-to-device (D2D) CSI, one can utilize the same method used in \cite{6195835}, where a novel approach to obtain the full CSI in a source-relay-destination topology considering an FD relaying was proposed. Particularly, based on \cite{6195835}, in our system model, we can assume that there is a control channel tunnelling from the source (BS), the relays (CCUs), and the destinations (CEUs) for channel estimation. Considering this discussion, we assumed that a perfect CSI is available at the centralized controller, i.e., the BBU pool. Moreover, as future work, we would like to extend our work by considering an imperfect and partial CSI assumption.
\subsection{Solution Approach}
Due to the intractability, we optimize the time slot allocation, $\theta \in  (0,1]$ through an exhaustive search method with step size 0.1. \footnote{Through experiments, we found that a smaller step size, such as $0.05$, did not significantly change the results. For example, with a step size of $0.05$, $\theta T$ becomes 0.15, and $(1-\theta) T$ becomes $0.85$ in the next iteration. Meanwhile, with a step size of $0.1$, $\theta T$ becomes $0.2$ and $(1-\theta) T$ becomes $0.8$ in the next iteration. The differences in results between these scenarios are negligible.} Specifically, we obtain the max-min rate value for each given $\theta$, and then select the value of $\theta$ that can achieve the maximum max-min rate. Thus, with a given value of $\theta$, we first introduce a slack variable $\beta$ representing the max-min rate. Then, we replace the objective function with $\beta$. Hence, $\mathcal{P}-\mathrm{HD}$ can be rewritten as,
\allowdisplaybreaks
\begin{subequations}
\label{prob: Main4}
\begin{align}
&\mathcal{P}-\mathrm{HD}: \max_{\substack{\textbf{P}^{\rm HD}, \textbf{v}^{\rm HD},\textbf{c}^{\rm HD},\\\boldsymbol{\beta}, \boldsymbol{\gamma}^{\rm HD}, \boldsymbol{\zeta}}} \ \; \quad \beta, \tag{P.2}   \\
&\textrm{s.t.}  \ \;  \mathrm{(\ref{P1_C1})-(\ref{P1_C6})},\nonumber\\
& C_m + \theta\log_2\left(1+\gamma_{k,p}^{[1], \rm HD}\right) \ge \beta, \quad \forall m\in \mathcal{M},\forall k\in \mathcal{K},\label{P2_C2}\\
& C_m + \theta\log_2\left(1+\gamma_{l,p}^{[1], \rm HD}\right) \ge \beta, \quad \forall m\in \mathcal{M}, \forall l\in \mathcal{L},\label{P2_C3}\\
& \frac{\textbf{h}_{k}\textbf{w}_{k,p}\textbf{w}_{k,p}^H\textbf{h}_{k}^H}{\textbf{h}_{k}\left(\sum_{j\in\mathcal{M}, j \ne k}\textbf{w}_{j,p}\textbf{W}_{j,p}^H\right)\textbf{h}_{k}^H + \sigma^2} \ge \gamma_{k,p}^{[1], \rm HD}, \forall k \in \mathcal{K},\label{P2_C4}\\
& 
{\frac{\textbf{h}_{l}\textbf{w}_{l,p}\textbf{w}_{l,p}^H\textbf{h}_{l}^H}{\textbf{h}_{l}\left(\sum_{j\in\mathcal{K}}\textbf{w}_{j,p}\textbf{w}_{j,p}^H\right)\textbf{h}_{l}^H + \sigma^2} \ge \gamma_{l,p}^{[1], \rm HD}, \forall l \in \mathcal{L}.\label{P2_C7}}
\end{align}
\end{subequations}
{It can be seen that \eqref{P2_C4} and \eqref{P2_C7} are still non-convex. To deal with \eqref{P2_C4} and \eqref{P2_C7}, we introduce slack variables $\zeta_{k,p}^{[1], \rm HD}, \zeta_{l,p}^{[1], \rm HD}$ $\forall k \in \mathcal{K}$ and $\forall l \in \mathcal{L}$ and transform \eqref{P2_C4} and \eqref{P2_C7} as follows,}
\allowdisplaybreaks
\begin{subequations}
\label{prob: Main4}
\begin{align}
&\frac{\textbf{h}_{k}\textbf{w}_{k,p}\textbf{w}_{k,p}^H\textbf{h}_{k}^H}{\zeta_{k,p}^{[1],\rm HD}} \ge \gamma_{k,p}^{[1], \rm HD},\quad \forall k \in \mathcal{K},\label{P2_C5}\\
& \zeta_{k,p}^{[1], \rm HD} \ge  \textbf{h}_{k}\left(\sum_{j\in\mathcal{M}, j \ne k}\textbf{w}_{j,p}\textbf{w}_{j,p}^H\right)\textbf{h}_{k}^H + \sigma^2, \quad \forall k \in \mathcal{K}, \label{P2_C6} \\
& {\frac{\textbf{h}_{l}\textbf{w}_{l,p}\textbf{w}_{l,p}^H\textbf{h}_{l}^H}{\zeta_{l,p}^{[1],\rm HD}} \ge \gamma_{l,p}^{[1], \rm HD}, \forall l \in \mathcal{L},}\label{P2_C8}\\
& {\zeta_{l,p}^{[1], \rm HD} \ge  \textbf{h}_{l}\left(\sum_{j\in\mathcal{M}}\textbf{w}_{j,p}\textbf{w}_{j,p}^H\right)\textbf{h}_{l}^H + \sigma^2,  \forall l \in \mathcal{L},} \label{P2_C9} 
\end{align}
\end{subequations}
{\eqref{P2_C5} and \eqref{P2_C8} are still non-convex.  To address this challenge, we approximate these constraints using the first-order Taylor approximation around a certain feasible point $\left({\textbf{w}}_{k,p}, {\textbf{w}}_{l,p},\zeta_{k,p}^{[1], \rm HD}, \zeta_{l,p}^{[1], \rm HD}\right)$. Hence, upper bounds for the quadratic-over-linear terms of these constraints can be derived by applying first-order Taylor approximations. Thus, if $\left({\textbf{w}}_{k,p}, {\textbf{w}}_{l,p}, \zeta_{k,p}^{[1], \rm HD}, \zeta_{l,p}^{[1], \rm HD}\right)$ is a feasible point of problem \eqref{P2_C5} and \eqref{P2_C8} , then}
 \begin{align}
&\frac{\textbf{h}_{k}{\textbf{w}}_{k}{\textbf{w}}_{k}^H\textbf{h}_{k}^H}{\zeta_{k,p}^{[1], \rm HD}} \geq
\frac{2{\tt{Re}}\left\{\left({{\textbf{w}}}_{k}^{(i)}\right)^H\textbf{h}_{k}^H\textbf{h}_{k}{{\textbf{w}}}_{k}^{(i)}\right\}}{\zeta_{k,p}^{[1], \rm HD}}  \notag\\
&
 -\frac{\textbf{h}_{k}{{\textbf{w}}}_{k}^{(i)}\left({{\textbf{w}}}_{k}^{(i)}\right)^H\textbf{h}_k^H}{\left(\zeta_{k,p}^{[1], \rm HD}\right)^2} \zeta_{k,p}^{[1], \rm HD} , \forall k \in \mathcal{K}, \label{Aprox1} 
\end{align}
\begin{align}
\frac{\textbf{h}_{l}{\textbf{w}}_{l,p}{\textbf{w}}_{l,p}^H\textbf{h}_{l}^H}{\zeta_{l,p}^{[1], \rm HD}} &\geq \frac{2{\tt{Re}}\left\{\left({{\textbf{w}}}_{l,p}^{(i)}\right)^H\textbf{h}_{l}^H\textbf{h}_{l}{{\textbf{w}}}_{l,p}^{(i)}\right\}}{\zeta_{l,p}^{[1], \rm HD}} \notag \\ &- \frac{\textbf{h}_k{{\textbf{w}}}_{l}^{(i)}\left({{\textbf{w}}}_{l}^{(i)}\right)^H\textbf{h}_{l}^H}{\left(\zeta_{l,p}^{[1], \rm HD}\right)^2} \zeta_{l,p}^{[1], \rm HD}, \forall l \in \mathcal{L}, \label{Aprox2}  
\end{align}
\eqref{P1_C1} can be handled in a similar way we handle the objective $\mathcal{P}-\mathrm{HD}$. At first, we introduce $\gamma_{k,c}^{[1], \rm HD}$,  $\gamma_{l,c}^{[1], \rm HD}$, $\gamma_{l,c}^{[2], \rm HD}$ and transform them as follows,
\allowdisplaybreaks
\begin{subequations}
\label{prob: Main44}
\begin{align}
&  \sum_{m\in\mathcal{M}}C_m \le \theta \log_2(1+ \gamma_{k,c}^{[1], \rm HD}), \quad \forall k \in \mathcal{K},\label{P2_C10}\\
& \frac{\textbf{h}_{k}\textbf{h}_c\textbf{w}_c^H\textbf{h}_{k}^H}{ \textbf{h}_{k}\left(\sum_{m\in\mathcal{M}}\textbf{w}_{m,p}\textbf{w}_{m,p}^H\right)\textbf{h}_{k}^H + \sigma^2} \ge \gamma_{k,c}^{[1], \rm HD}, \quad \forall k \in \mathcal{K}, \label{P2_C11}\\
&  \sum_{m\in\mathcal{M}}C_m \le \theta \log_2(1+ \gamma_{l,c}^{[1], \rm HD}) + \notag\\
&(1-\theta) \log_2(1+\gamma_{l,c}^{[2], \rm HD}),\forall l \in \mathcal{L},\label{P2_C14}\\
&\frac{\textbf{h}_{l}\textbf{w}_c\textbf{w}_c^H\textbf{h}_{l}^H}{ \textbf{h}_{l}\left(\sum_{m\in\mathcal{M}}\textbf{w}_{m,p}\textbf{w}_{m,p}^H\right)\textbf{h}_{l}^H + \sigma^2} \ge \gamma_{l,c}^{[1], \rm HD},  \quad \forall l  \in \mathcal{L},\label{P2_C15} \\
& \frac{|\textbf{h}_{l,d}^H \textbf{v}^{\rm{HD}}|^2}{\sigma^2}  \ge \gamma_{l,c}^{[2], \rm HD},\label{P2_C18}
\end{align}
\end{subequations}
{We can handle \eqref{P2_C11} and \eqref{P2_C15} similar manner as we solve \eqref{Aprox1} and \eqref{Aprox2}. Meanwhile, \eqref{P2_C18} is non-convex and to handle \eqref{P2_C18}, we invoke first-order Taylor approximation and approximate as follows,}
\begin{align}
&\frac{|\textbf{h}_{l,d}^H \textbf{v}^{\rm{HD}}|^2}{\sigma^2} \geq  \frac{\textbf{h}_{l,d}^H \textbf{v}^{{\rm{HD}},(i)} \left(\textbf{v}^{{\rm{HD}},(i)}\right)^H \textbf{h}_{l,d}}{\sigma^2}\notag\\
     &+ \frac{2 {\tt{Re}}\left\{\left(\textbf{v}^{{\rm{HD}},(i)}\right)^H \textbf{h}_{l,d} \textbf{h}_{l,d}^H \left( \textbf{v}^{{\rm{HD}}} - \textbf{v}^{{\rm{HD}},(i)} \right)\right\}}{\sigma^2}. \label{Aprox4} 
\end{align}
Moving onto \eqref{P1_C2}, utilizing first-order Taylor approximation, we can approximate \eqref{P1_C2} as
\begin{equation}
    \sum_{m\in\mathcal{M}} (C_m + \theta g_m) - F_{n_b}^{\rm Fh} \le 0, \quad \forall n_b \in \mathcal{N}_b, \label{Approx51}
\end{equation}
\begin{align}
    &\log_2(1+\gamma_{m,p}^{[1], \rm HD, (i)})+\notag\\
    &\frac{1}{(1+\gamma_{m,p}^{[1],\rm HD, (i)})\ln(2)}(\gamma_{m,p}^{[1],\rm HD}-\gamma_{m,p}^{[1], \rm HD, (i)}) \le g_m, \quad \forall m \in \mathcal{M}, \label{Approx61} 
\end{align}
Finally, using the above approximations, $\mathcal{P}-\mathrm{HD}$ can be transformed as
\allowdisplaybreaks
\begin{subequations}
\label{prob: Main4}
\begin{align}
&\mathcal{\tilde{P}}-\mathrm{HD}: \max_{\substack{\textbf{P}^{\rm HD}, \textbf{v}^{\rm HD}, \textbf{c}^{\rm HD}, \\ {\beta},  \boldsymbol{\gamma}^{\rm HD}, \boldsymbol{\zeta}}} \ \; \beta, \tag{P.3}  \\
&\textrm{s.t.}  \ \; \frac{\textbf{h}_{k}{\textbf{w}}_c{\textbf{w}}_c^H\textbf{h}_{k}^H}{\zeta_{k,c}^{[1], \rm HD}} \geq \frac{2{\tt{Re}}\left\{\left({{\textbf{w}}}_c^{(i)}\right)^H\textbf{h}_{k}^H\textbf{h}_{k}{{\textbf{w}}}_c^{(i)}\right\}}{\zeta_{k,c}^{[1], \rm HD}} - \notag \\
&\frac{\textbf{h}_{k}{{\textbf{w}}}_c^{(i)}\left({{\textbf{w}}}_{k}^{(i)}\right)^H\textbf{h}_{k}^H}{\left(\zeta_{k,c}^{[1], \rm HD}\right)^2} \zeta_{k,c}^{[1], \rm HD} , \forall k \in \mathcal{K}, \label{Aprox31} \\
&\frac{\textbf{h}_{l}{\textbf{w}}_c{\textbf{w}}_c^H\textbf{h}_{l}^H}{\zeta_{l,c}^{[1], \rm HD}} \geq \frac{2{\tt{Re}}\left\{\left({{\textbf{w}}}_c^{(i)}\right)^H\textbf{h}_{l}^H\textbf{h}_{l}{{\textbf{w}}}_c^{(i)}\right\}}{\zeta_{l,c}^{[1], \rm HD}} - \\
&\frac{\textbf{h}_{l}{{\textbf{w}}}_c^{(i)}\left({{\textbf{w}}}_{c}^{(i)}\right)^H\textbf{h}_{l}^H}{\left(\zeta_{l,c}^{[1], \rm HD}\right)^2} \zeta_{l,c}^{[1], \rm HD}, \forall l \in \mathcal{L},\label{Aprox45}  \\
& \zeta_{k,c}^{[1],\rm HD} \ge \textbf{h}_{k}\left(\sum_{m\in\mathcal{M}}\textbf{w}_{m,p}\textbf{w}_{m,p}^H\right)\textbf{h}_{k}^H + \sigma^2, \quad \forall k \in \mathcal{K}, \label{P2_C131}\\
&\zeta_{l,c}^{[1], \rm HD} \ge \textbf{h}_{l}\left(\sum_{m\in\mathcal{M}}\textbf{w}_{m,p}\textbf{w}_{m,p}^H\right)\textbf{h}_{l}^H + \sigma^2, \forall l \in \mathcal{L}, \label{P2_C17}\\
&\eqref{P1_C3} - \eqref{P1_C6}, \eqref{P2_C2}, \eqref{P2_C3},\eqref{P2_C6}, \eqref{P2_C9}, \notag \\
&\eqref{Aprox2},\eqref{P2_C10}, \eqref{P2_C14}, \eqref{Aprox4},\eqref{Approx61}.\nonumber
\end{align} 
\end{subequations}
One can see that all the constraints in $\mathcal{\tilde{P}}-\rm HD$ are transformed into convex constraints, which can be solved using any convex optimization toolbox such as CVX. The SCA-based overall algorithm for HD CoMP-C-RSMA is provided in Algorithm 1. In the following, we discuss the initialization process, the convergence analysis, and the computational complexity of the proposed solution approach.
\par \textit{Initialization}: The beamforming vectors $\textbf{P}^{[0]}$ are initialized by obtaining the feasible beamforming vectors that satisfy the constraints (\ref{P1_C1})-(\ref{P1_C6}) \cite{9627180}. After obtaining the beamformers, we can start Algorithm 1 whereas, $\boldsymbol{\zeta}_{k,p}^{[1], \rm HD}$,$\boldsymbol{\zeta}_{l,p}^{[1], \rm HD}$, $\boldsymbol{\zeta}_{k,c}^{[1], \rm HD}$, $\boldsymbol{\zeta}_{l,c}^{[1], \rm HD}$, $\boldsymbol{\gamma}_{m,p}^{[1], \rm HD}$ are initialized by replacing inequalities in \eqref{P2_C6}, \eqref{P2_C9}, \eqref{P2_C131}, \eqref{P2_C17}, \eqref{P2_C4} and \eqref{P2_C7} respectively \cite{8491100}. Meanwhile, $\beta$ is initialized considering an initial maximum rate, and $\textbf{v}^{\rm HD}$ is initialized randomly by satisfying the power budget constraint of UEs \cite{8491100}.
\par \textit{Convergence}:
{Following \cite{8648507,beck2010sequential}, the convergence of our proposed SCA-based algorithm can be measured. As presented in Algorithm 1, it can be seen that the optimization parameters at $n$-th iteration are updated based on the solution that can be obtained by solving the optimization problem in Eqn 29. Three conditions are required to be satisfied in our convergence analysis. First, the optimization problem needs to be initialized appropriately with initial parameters ($\rm \textbf{P}^0$,  $\beta^{0}, \boldsymbol{\gamma}^{0,\rm HD}_{m,p}, \boldsymbol{\zeta}^{0}$) which ensures the feasibility of the problem at each iteration and provides a feasible solution to update the parameters for next iteration. Second, the optimization variable $\beta$ is linear, hence this variable is always non-decreasing and holds $\beta^n \ge \beta^{n-1}$. Last but not the least, as shown in constraints (20d)-(20g), the beamforming vectors and relaying power are upper bounded by the BS transmission power and CCU relaying power such that $P_{n_b}^{\rm \max} < \infty$ and $P_{k}^{\rm{tot}} < \infty$ which implies
that $\beta$ is upper bounded, as well. 
Meeting these three conditions guarantees that the developed SCA technique will converge to a solution within a finite number of iterations.}
\par {\textit{Computational Complexity analysis}:
In order to measure the computational complexity of Algorithm 1, we need to analyze the complexity of the exhaustive search and SCA process. It should be noted that we conduct an exhaustive search within the interval  (0, 1] with a step size of 0.1 to find the optimal time slot duration $\theta$ that maximizes the minimum rate, and each iteration $\mathcal{\tilde{P}}-\mathrm{HD}$ is solved. The complexity burden of $\mathcal{\tilde{P}}-\mathrm{HD}$ arises from the computational task of solving the SCA process. That being said, $\mathcal{\tilde{P}}-\mathrm{HD}$ is a second-order cone problem and it has the complexity
of $O(KN_t)^{3.5}$ \cite{9123680}. The
total number of iterations required for the convergence is approximated as $O(\log(\epsilon_1)^{-1})$. Hence, the computational complexity of Algorithm 1 can be calculated as $O(J_1\log(\epsilon_1)^{-1}(KN_t)^{3.5})$ where $J_1$ represents the number of iterations required for exhaustive search.}
\begin{algorithm}[!t]
\caption{SCA-based algorithm for HD CoMP-enabled C-RSMA Networks}\label{alg:one}
\KwIn{Time slot allocation $\theta$, tolerance $\epsilon_1$, number of iterations $J$}, $\theta_{min}=0.1$, $\theta_{max}=1$\\
\textbf{Initialize}: \rm $\textbf{P}^0$,  $\beta^{0}$, $\boldsymbol{\gamma}^{[1],\rm HD}_{m,p}$, $\boldsymbol{\zeta}^{0}$, $\textbf{v}^{\rm HD}$\;
$n=0$\;
\For {$\theta =0.1:0.1:1$}{
\While{$|\beta^n$ - $\beta^{n-1}| > \epsilon_1$ or $n \le J $}{
$n=n+1$\;  
solve $\mathcal{\tilde{P}}-\mathrm{HD}$ using  $\textbf{P}^{n-1},
 \beta^{n-1}$, $\boldsymbol{\gamma}^{[1],\rm HD^{n-1}}_{m,p}$, $\boldsymbol{\zeta}^{n-1}$,  $\textbf{v}^{\rm HD^{n-1}}$\;
  Find the optimization variables $\textbf{P}^*,
\beta^*$,$\boldsymbol{\gamma}^{[1],\rm HD^*}_{m,p}$, $\boldsymbol{\zeta}^*$, $\textbf{v}^{\rm HD^{*}}$;\\  
  Update $\textbf{P}^n$ $\leftarrow \textbf{P}^*$, $\beta^n \leftarrow \beta^*$, $\boldsymbol{\gamma}^{[1],\rm HD^{n}}_{m,p} \leftarrow \boldsymbol{\gamma}^{[1],\rm HD^{*}}_{m,p}$,$ \boldsymbol{\zeta}^n \leftarrow \boldsymbol{\zeta}^*$, $\textbf{v}^{\rm HD^{n}} \leftarrow \textbf{v}^{\rm HD^{*}}$;
}
$R_{\theta}= \beta^*$;
}
$R_{opt}=\max(R_{\theta})$;
\end{algorithm}
\section{Coordinated FD C-RSMA: Problem Formulation and Solution Approach}
\subsection{Problem Formulation}
In this section, we investigate both the problem formulation and the proposed solution approach for CoMP-assisted FD C-RSMA-enabled wireless networks. More precisely, we investigate the joint optimization of beamforming vectors at the BSs $\left(\textbf{P}^{\rm{FD}}\right), \forall n_b\in \mathcal{N}_b$, the distributed beamforming at the CCUs $\left(\textbf{v}^{\rm{FD}}\right)$, and the common stream split $\left(\textbf{c}^{\rm{FD}}= \{C_1^{\rm{FD}}, \dots, C_{M}^{\rm{FD}}\}\right) $. Accordingly, we formulate this framework as an optimization problem aiming to maximize the minimum user data rate. Thus, the max-min rate optimization problem for the proposed coordinated FD C-RSMA system can be formulated as 
\allowdisplaybreaks
\begingroup
\begin{subequations}
\label{prob: Main2}
\begin{align}
\mathcal{P}-\mathrm{FD}: &\quad \max_{\substack{\textbf{P}^{\rm{FD}}, \label{P2} \textbf{v}^{\rm{FD}}\\\textbf{c}^{\rm{FD}}}} \min \mathcal{R}_m^{\rm{FD}} \qquad m \in \mathcal{M},  \\
\text{s.t.}\,\, & \quad \,\,\, \sum_{m\in\mathcal{M}} C_m^{\rm{FD}} \leq R_c^{\rm{FD}}, \label{P3_C1}\\
&\quad \,\,\, \mathrm{tr}\left(\textbf{P}_{n_b}^{\rm{FD}}(\textbf{P}_{n_b}^{\rm{FD}})^H\right) \leq P_{n_b}^{\rm \max}, && n_b \in \mathcal{N}_b,\label{P3_C2}\\
&\quad \,\,\, \sum_{m\in\mathcal{M}} \mathcal{R}_m^{\rm{FD}} \leq F^{\rm{Fh}}_{n_b}, && n_b \in \mathcal{N}_b, \label{P3_C3}\\
&\quad \,\,\, |v_{k}^{\rm{FD}}|^2\leq P_{k}^{\rm{tot}}, && k \in \mathcal{K},\label{P3_C4}\\
&\quad \,\,\, C_m^{\rm{FD}} \geq 0, && m \in \mathcal{M}. \label{P3_C5}
\end{align}
\end{subequations}
\endgroup
Similar to $\mathcal{P}-\mathrm{HD}$, problem  $\mathcal{P}-\mathrm{FD}$ is hard to solve with common standard optimization techniques due to the non-convex objective and constraints. To tackle this challenge, surrogate upper-bound functions are derived for all non-convex terms. As a result, an approximation for the non-convex terms to a convex ones can be obtained. Afterward, these approximations are iteratively updated until convergence by applying the SCA approach. The SCA-based overall algorithm for FD CoMP C-RSMA is provided in Algorithm 2.
 Note that we can set the initial feasible points for Algorithm 2 as those of Algorithm 1. Moreover, both processes follow the SCA-based approach. Hence, the initialization, convergence, and complexity of Algorithm 2 are similar to Algorithm 1, and therefore omitted for brevity. {The SCA-based solution approach for Algorithm 2 has been moved to Appendix.}
\begin{algorithm}[!t]
\caption{SCA-based algorithm for FD CoMP-CRSMA}\label{alg:one}
\KwIn{tolerance $\epsilon_2$, Number of iterations $J$}
\textbf{Initialize}: \rm $\textbf{P}^0$,  $\alpha^0$, $\boldsymbol{\gamma}_{m,p}^{0, \rm FD}$, $\boldsymbol{\mu}_{m,c}^{0,\rm FD}$, $\boldsymbol{\mu}_{m,p}^{0, \rm FD}$\;
$n=0$\;
\While{$|\alpha^n$ - $\alpha^{n-1}| > \epsilon_2$ or $n \le J$}{
$n=n+1$\;  
solve $\mathcal{\tilde{P}}-\mathrm{FD}$ using  $\textbf{P}^{n-1},
 \alpha^{n-1}$, $\boldsymbol{\gamma}_{m,p}^{n-1, \rm FD}$, $\boldsymbol{\mu}_{m,c}^{n-1}$, $\boldsymbol{\mu}_{m,p}^{n-1}$\;
  Find the optimization variables $\textbf{P}^*,
\alpha^*$, $\boldsymbol{\gamma}_{m,p}^{*, \rm FD}$, $\boldsymbol{\mu}_{m,c}^*$, $\boldsymbol{\mu}_{m,p}^*$;\\  
  Update $\textbf{P}^n$ $\leftarrow \textbf{P}^*$, $\boldsymbol{\gamma}_{m,p}^{n, \rm FD} \leftarrow \boldsymbol{\gamma}_{m,p}^{*, \rm FD}$, $ \boldsymbol{\mu}_{m,c}^{n, \rm FD} \leftarrow \boldsymbol{\mu}_{m,c}^{*,\rm FD}$, $ \boldsymbol{\mu}_{m,p}^{n,FD} \leftarrow \boldsymbol{\mu}_{m,p}^{*,\rm FD}$;
}
\end{algorithm}
\section{Simulation results \& discussions}
\begin{table}
\centering
\caption{{Simulation parameters}}
\begin{tabular}{ |l|c|c| }
 \hline
  \textbf{Parameter} & \textbf{Symbol} & \textbf{Value} \\
 \hline
 Power budget of BSs & $P_{n_b}^{\rm \max}$  & 30 dBm  \\
 \hline
 Power budget of CCUs & $P_{k}^{\rm tot}$ &   20 dBm \\
 \hline
Number of CCUs & $K$ & 2 \\
 \hline
Number of BSs & $N$ & 2 \\
 \hline
 Number of CEUs& $L$ & 2  \\
 \hline

 Fronthaul capacity & $F_{n_b}^{\rm Fh}$& 20 bps/Hz \\
 \hline
 Self-interference & $\rm SI$& -10 dB\\
 \hline
 Number of antennas at each BS &$N_t$& 4\\
 \hline
 {\makecell[l]{Channel disparity between \\ CCU-1, and BS-1 \\ and CCU-2 and BS-2 }}
 &$\delta_s$ & 1\\
 \hline
  {\makecell[l]{Channel disparity between \\ CCU-1, and BS-2, \\ CCU-2 and BS-1}} &$\delta_{s_w}$ & 0.1\\
 \hline
 {\makecell[l]{Channel disparity between \\ CEU-1, and BS-1, \\
 CEU-1 and BS-2}} &$\delta_w(1)$ & 0.3\\
 \hline
{\makecell[l]{Channel disparity between\\  CEU-2 and BS-1, \\
CEU-2 and BS-2}} &$\delta_w(2)$ & 0.4\\
\hline
{\makecell[l]{Channel disparity between \\ CCU-1 and CEU-1, \\CCU-2 and CEU-2}}&$\delta_d (1)$ & 1\\
\hline
{\makecell[l]{Channel disparity between \\CCU-1 and CEU-2,\\
CCU-2 and CEU-1 }}&$\delta_d (2)$ & 1 and 0.8\\
\hline
\end{tabular}
\end{table}
In this section, the efficacy of the proposed system model
and the solution approach is evaluated. Unless otherwise specified, the number of cells $= 2$, where each cell contains one BS and one CCU. { The parameters of wireless channels are assumed to be as follows: $\lambda_s = 12$ dB, $\lambda_w = 3$ dB, $\lambda_d = 9$ dB. We utilized different channel disparities between BS and CCU, BS and CEU, and CCU and CEU, which are denoted by $\delta_s$, $\delta_w$, and $\delta_d$, respectively.} The details of the simulation parameters are provided in Table II.  {Unless otherwise specified, we adopt the parameters in Table II throughout our simulation. The simulations are performed on a desktop with an Intel Core 12th Gen i7-12700K 3.60 GHz CPU and 32GB RAM and MATLAB R2020b is used for the proposed algorithm.} We compare our proposed scheme
to four benchmark schemes under different system parameters such as different transmission power levels for the BSs and UEs, various channel strength disparities, different SI channels, different numbers of BSs, and different ranges of fronthaul capacities. These adopted benchmark schemes are,
\begin{itemize}
\item \textbf{CoMP RSMA scheme} \cite{8756668}: This scheme is similar to the proposed scheme but without the assistance of user relaying.
\item \textbf{CoMP FD C-NOMA scheme} \cite{9737471}:  The system model is similar to our proposed system model. However, the used MA technique is different, NOMA is adopted.
\item \textbf{CoMP HD C-NOMA scheme} \cite{9737471}: This scheme is similar to the previous scheme but in HD relaying mode.
\item \textbf{CoMP NOMA scheme} \cite{8352643}: This scheme considered the amalgamation between the JT-CoMP and NOMA schemes without cooperation between the cellular users.  
\end{itemize}
\subsection{Impact of channel disparity vs max-min achievable rate}
\begin{figure}
    \centering
\includegraphics[scale=0.5]{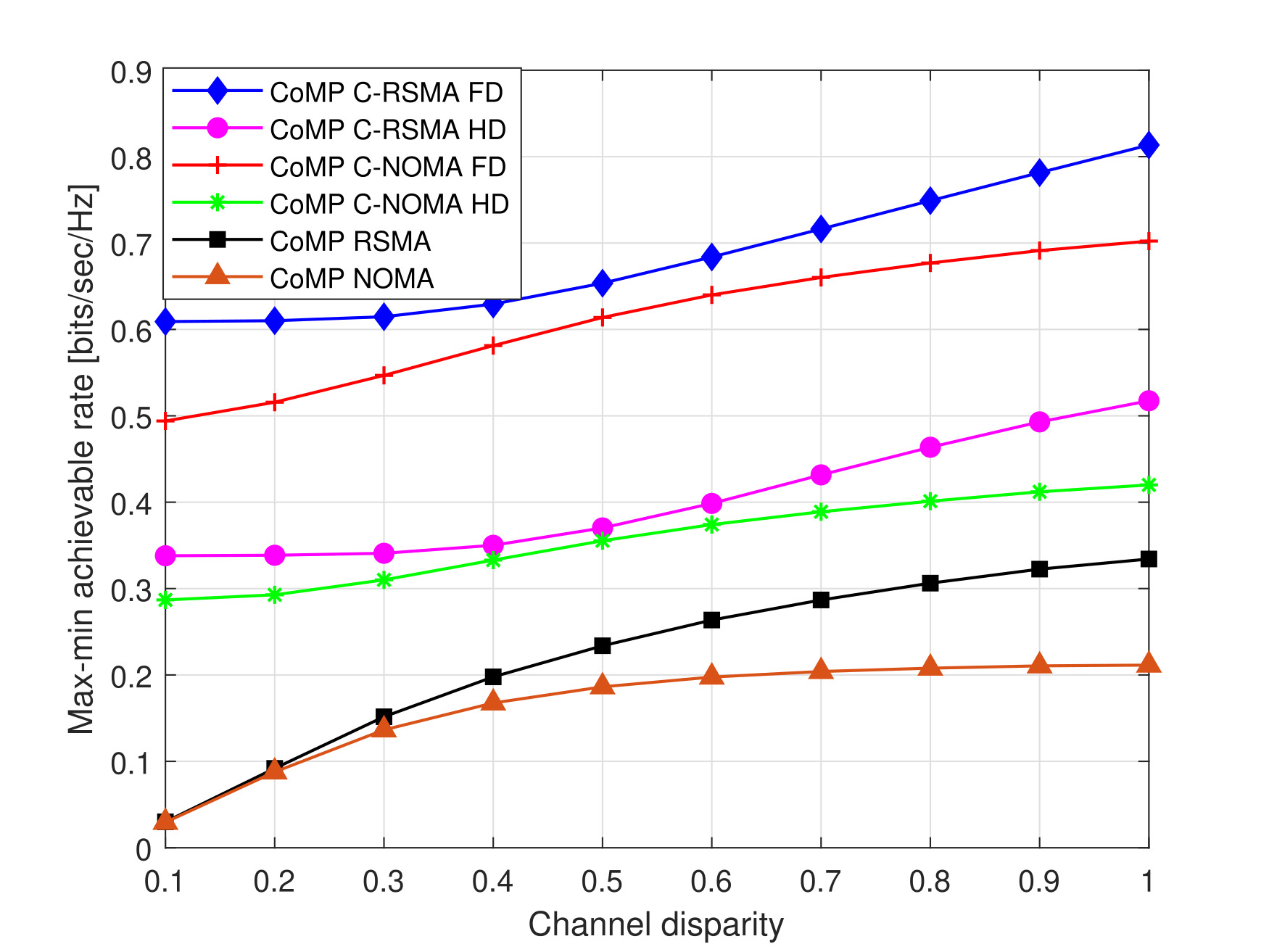}
    \caption{{Max-min achievable rate vs channel disparity of $\delta_w (1)$, when $P_{n_b}^{\rm \max}$ = 20 dBm and $\delta_w(2)$ = 0.4}}
\label{channel_disparity_1}
\end{figure}

{Fig. \ref{channel_disparity_1} evaluates the effect of channel disparity at the CEUs in our proposed scheme and compares it with other baseline schemes. We vary $\delta_w$ of CEU-1 from 0.1 to 1 where 0.1 represents high channel disparity and 1 represents no channel disparity. It can be seen from the figure that when we move from 0.1 towards 1, the max-min achievable rate begins to increase for all schemes. This is because as channel disparity decreases, the channel gain between the CEU-1 and BSs becomes better, resulting in an improvement in the achievable data rate. However, it is noticeable that C-RSMA and RSMA achieve higher performance gains than C-NOMA and NOMA. This is because NOMA-based schemes perform well in high channel strength disparity and hence achieve similar performance as RSMA-based schemes; meanwhile, RSMA-based schemes can perform well in various channel strength disparities.}
\begin{figure}
    \centering
\includegraphics[scale=0.5]{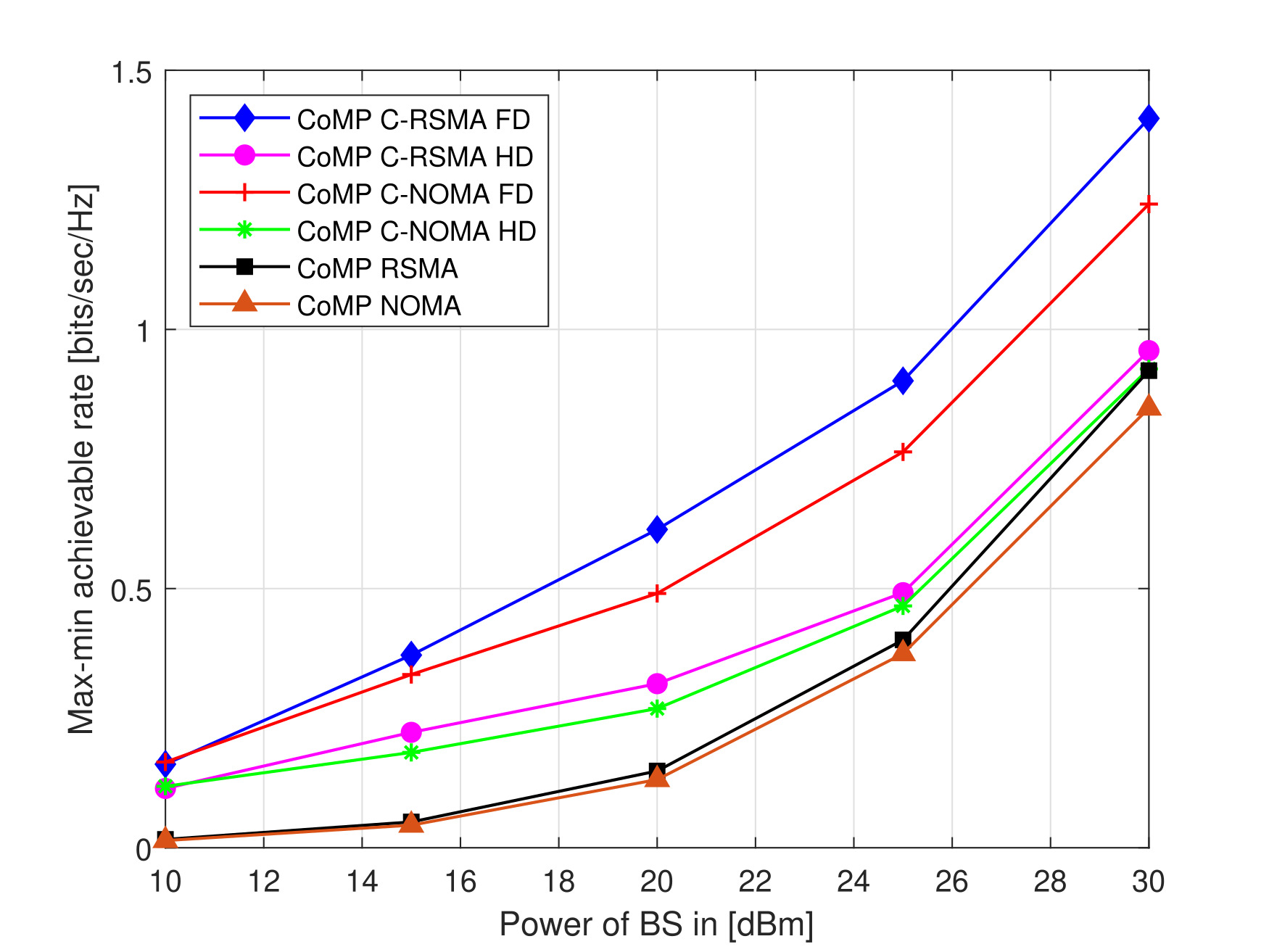}
    \caption{{Max-min achievable rate vs power of BS, when $P_{k}^{\rm tot}$ = 20 dBm and $\delta_d(1)$ = 1, $\delta_d(2)$ = 1.}}
\label{pow_BS}
\end{figure}
\begin{figure}
    \centering
\includegraphics[scale=0.5]{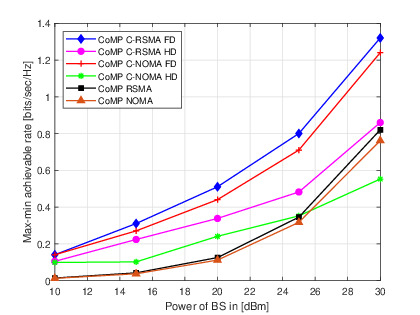}
    \caption{{Max-min achievable rate vs power of BS, when $P_{k}^{\rm tot}$ = 20 dBm and $\delta_d(1)$ = 1, $\delta_d(2)$ = 0.8}}
\label{pow_BS2}
\end{figure}
\subsection{Impact of the power budget of BSs and inter-user interference on max-min achievable rate}
\par {In Fig. \ref{pow_BS}, we evaluate the effect of the BS power budget on the performance of the four baseline schemes compared to the achievable performance of the proposed scheme. As we increase the power budget of BS from 10 dBm to 30 dBm, the rates of all schemes start to increase. This is because when the BSs possess more power budget, it can overcome the bad channel gains for the CEUs. This eventually results in a higher max-min achievable rate.  However, CoMP-based FD C-RSMA achieves a superior performance gain over all other schemes. It can also be seen from this figure that, CoMP-based HD C-RSMA and CoMP-based HD C-NOMA achieve higher performance than CoMP RSMA and CoMP NOMA in low power of BS. However,  CoMP RSMA and CoMP NOMA tend to increase toward CoMP-based HD C-RSMA and CoMP-based HD-C-NOMA, in high power of BS. It is because CoMP-based HD C-RSMA and CoMP-based HD-C-NOMA suffer from pre-log penalty which limits them from achieving higher rates.}
{
Fig. \ref{pow_BS} shows the max-min achievable rate when low channel disparity exists between each CCU and each CEU, the value of channel disparity is 1. In this scenario, for instance, when the transmit power at the BS is set to $20$ dBm, the gain achieved by CoMP C-RSMA FD is approximately $25\%$ higher than that of CoMP C-NOMA FD. Conversely, Fig. \ref{pow_BS2} illustrates the max-min achievable rate when a moderate channel disparity of $0.8$ exists between each CCU and CEU. In this particular case, CoMP C-RSMA FD demonstrates around a $14\%$ gain over CoMP C-NOMA FD at $P_{{k_b}}^{\rm max} = 20$ dBm. Moreover, CoMP C-RSMA HD also shows significant improvement over the CoMP C-NOMA HD technique. This is because the C-RSMA-based techniques adopt a 1-layer C-RSMA technique where the common stream is relayed during the cooperative phase to all the CEUs without being affected by inter-user interference. On the other hand, the C-NOMA-based technique suffers from inter-user interference during the cooperative phase, which negatively impacts the max-min achievable rate. These results validate and show the superiority of 1-layer C-RSMA over C-NOMA-based techniques.}
\subsection{Impact of the power budget of CCUs vs max-min achievable rate}
\par
\begin{figure}
\centering\includegraphics[scale=0.5]{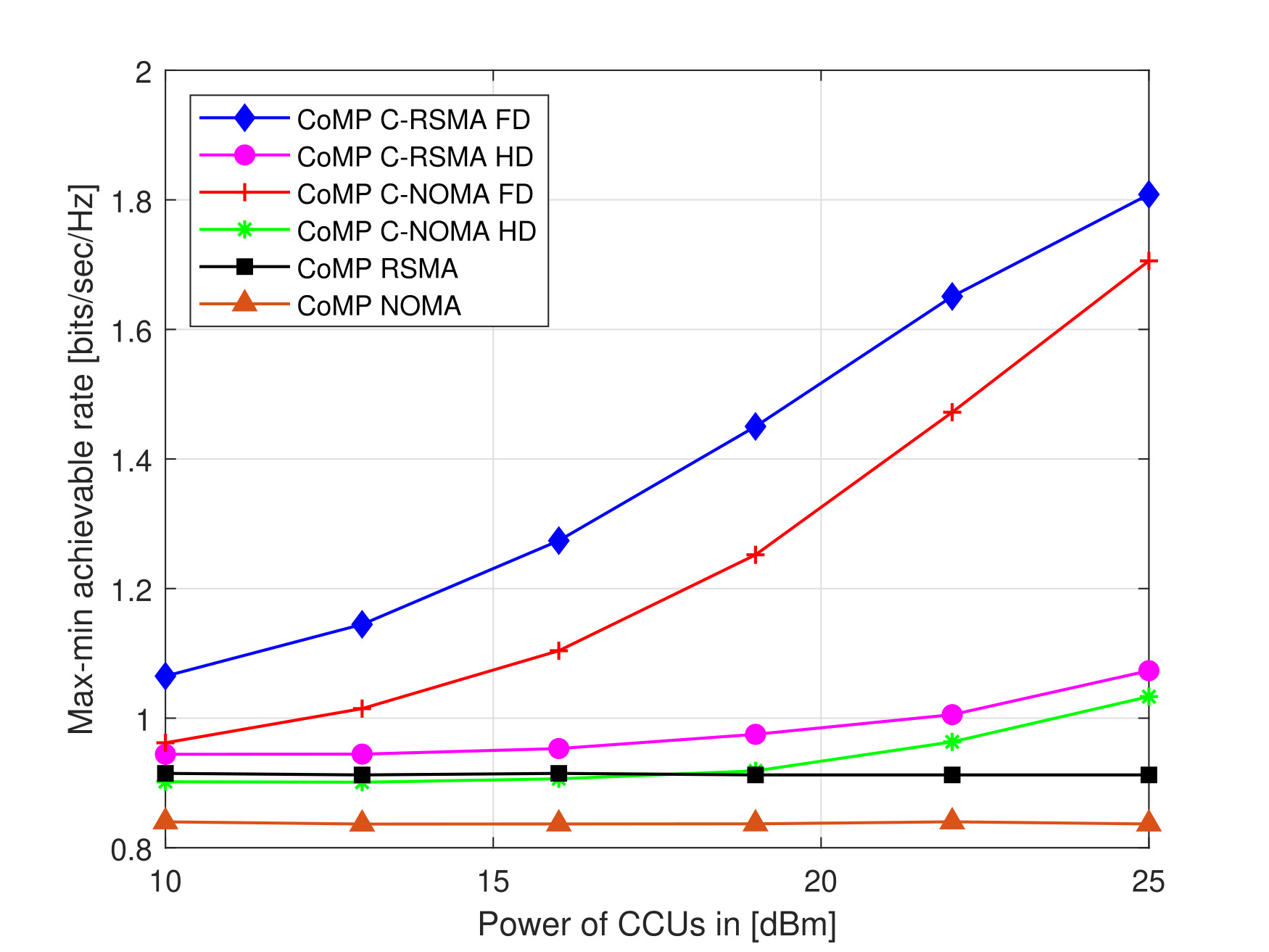}    \caption{{Max-min achievable rate vs  Power budget of CCUs.}}
\label{UE}
\end{figure}

 {In Fig. \ref{UE}, we vary the CCUs transmit power budget to evaluate the effect of user cooperation on the performance of the cooperative-based schemes. It is observed that as we increase the power budget of CCUs, the max-min achievable rate increases for all cooperative-based schemes. This is because when we increase the power budget of CCUs, these CCUs can spend more power to transmit the common stream to the CEUs, resulting in improved signal quality at the CEUs. Even though CoMP-based FD C-RSMA and CoMP-based FD C-NOMA can achieve a very high increase in rate, CoMP-based HD C-RSMA and COMP-based HD C-NOMA improve very slowly. It is because they suffer from pre-log penalty and even high CCU power cannot overcome it.  However, CoMP-based HD C-RSMA still shows better performance than CoMP-based HD C-NOMA, CoMP-based NOMA and CoMP-based RSMA. It is noteworthy that CoMP-based RSMA and CoMP-based NOMA are not affected by the different power budgets of CCUs as they do not perform cooperation.}
\subsection{Impact of the SI channel vs max-min achievable rate}
\par
\begin{figure}
\centering\includegraphics[scale=.5]{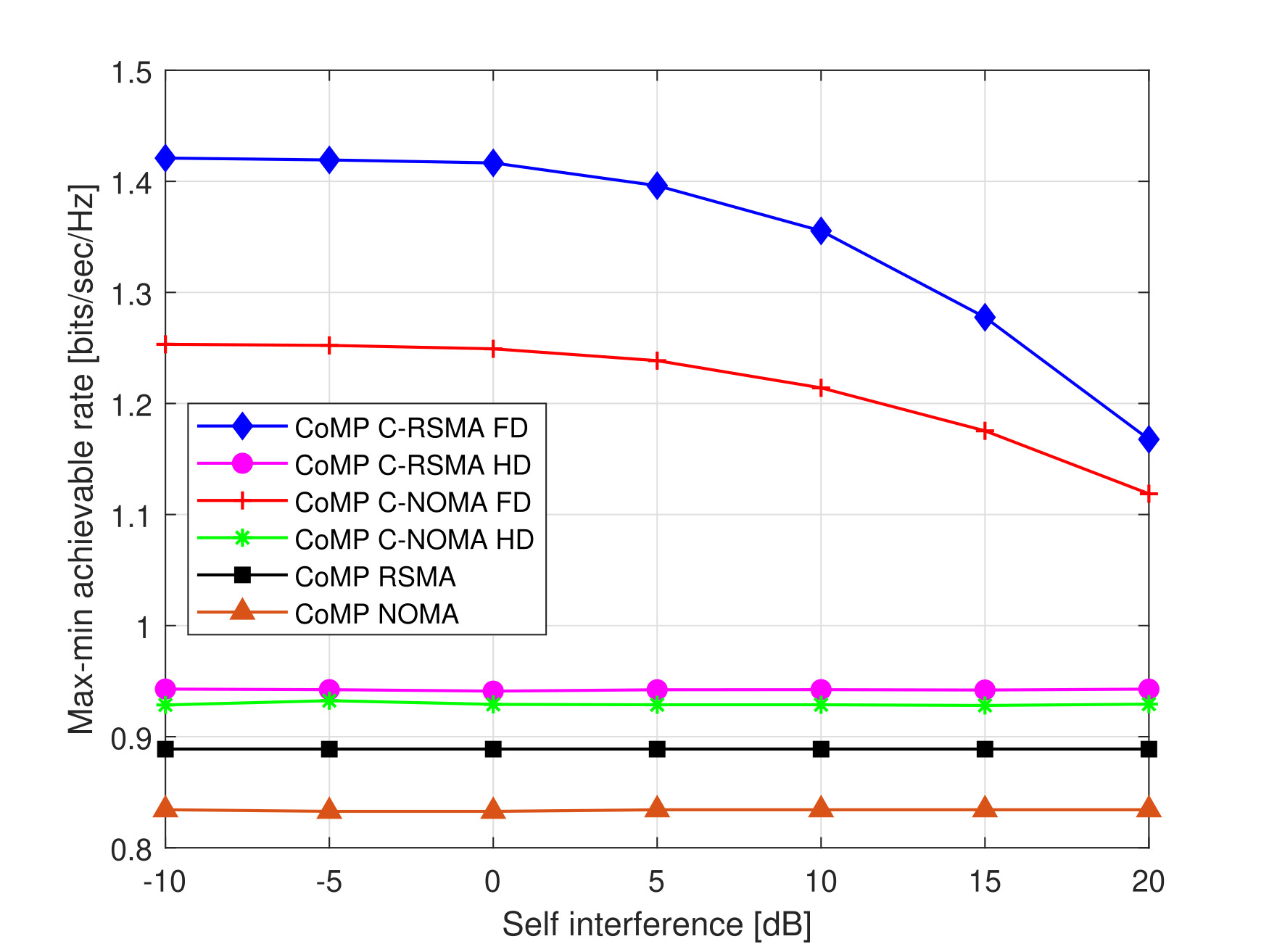}    \caption{{Max-min achievable rate vs SI channel.}}
\label{SI}
\end{figure}

{Fig. \ref{SI} shows the effect of the SI channel on the proposed scheme and the baseline schemes. It is observed that in low SI value, CoMP-based FD C-RSMA has a superior performance over the other five schemes. However, as we increase the value of the SI channel, the rate starts to decrease. This is because as we increase the SI channel, the BS tends to transmit with less power to overcome the negative impacts of the SI channel. However, even in a high SI channel of $20$ dB, CoMP-based FD C-RSMA achieves high-performance gain over other schemes, meanwhile, CoMP-based FD C-NOMA scheme tends to achieve lower performance gains compared to CoMP-based HD C-RSMA scheme in a high SI channel. It should be noted that CoMP-based HD C-RSMA, CoMP-based HD C-NOMA, CoMP-based NOMA and CoMP-based RSMA are not affected by the SI channel, hence their rate remains steady.}

\begin{figure}
\centering\includegraphics[scale=0.5]{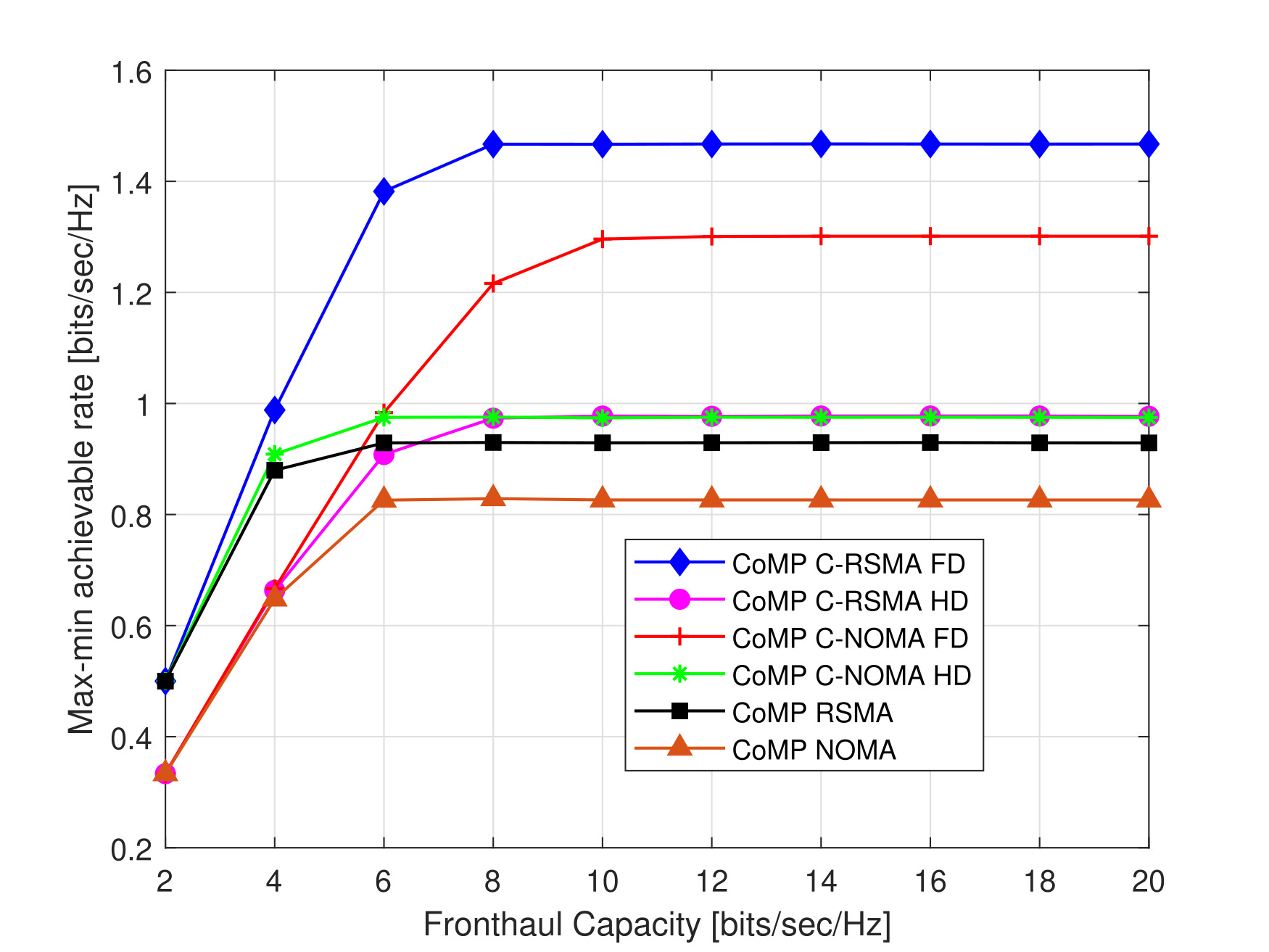}    \caption{{Max-min achievable rate vs fronthaul capacity}}
\label{fronthaul}
\end{figure}
\subsection{Impact of the fronthaul capacity vs max-min achievable rate}
\par {One can see that the FD schemes achieve higher performance gains compared to the HD schemes. Hence, in Fig \ref{fronthaul}, we evaluate the impact of fronthaul capacity on the FD schemes. It can be seen from the figure that when we increase the fronthaul capacity from 2 to 8 bps, the max-min achievable rate of the FD C-RSMA scheme increases accordingly. However, when we increase the fronthaul capacity from 8 bps to 20 bps, the max-min achievable rate stops increasing further and remains steady. Similarly, the rate of FD C-NOMA schemes increases until 10 bps and then remains steady. This occurs because, for a given power budget of BSs and CCUs, the system can achieve a max-min rate of around 1.5 bps/Hz for C-RSMA FD and around 1.3 bps/Hz for FD C-NOMA. For this reason even though the fronthaul capacity increases, the BS and CCUs cannot exert more energy to achieve a higher rate due to power budget constraints, and hence the max-min data rate remains steady.}
 \section{Conclusion and Future Work}
 With the goal of enhancing the performance of cellular systems in terms of max-min achievable rate, we investigated the joint beamforming at BSs, common stream rate allocation, relaying power allocation, and time slot duration in a CoMP-assisted C-RSMA network considering both HD and FD modes of relaying, which was formulated as a QCQP optimization problem. To tackle the non-convexity of the formulated problem, we invoked an SCA-based low-complexity algorithm that iteratively solved the formulated problem. By harnessing the flexible interference management capabilities of RSMA and leveraging the advantages of cooperative communications, our proposed C-RSMA-based schemes demonstrated significant performance improvements compared to baseline schemes. This is validated through extensive simulation results where we showed that our proposed CoMP-assisted FD C-RSMA can outperform other baseline schemes such as CoMP-assisted FD C-NOMA, RSMA, and NOMA. In addition, our proposed CoMP-assisted HD C-RSMA can outperform CoMP-assisted HD C-NOMA, RSMA, and NOMA. 
\section{Appendix}
\subsection{Solution Approach}
We solve $\mathcal{P}-\mathrm{FD}$ utilizing SCA similar manner as $\mathcal{P}-\mathrm{HD}$. Hence, we omit the details for brevity. After introducing slack variables and performing all the approximations, $\mathcal{P}-\mathrm{FD}$ becomes, 
\allowdisplaybreaks
\begingroup
\begin{subequations}
\label{prob: Main3}
\begin{align}
&\mathcal{\tilde{P}}-\mathrm{FD}: \quad \max_{\substack{\textbf{P}^{\rm{FD}}, \textbf{v}^{\rm{FD}}\textbf{c}^{\rm{FD}},\\ \alpha, \boldsymbol{\gamma}^{\rm FD}, \boldsymbol{\mu}}} \alpha,  \\
&\text{s.t.}\,\, \qquad\eqref{P3_C2}, \eqref{P3_C4}, \eqref{P3_C5},\notag \\
&\quad \,\,\, \sum_{m\in\mathcal{M}} C_m^{\rm{FD}} \leq \log_2\left(1 + \gamma^{\rm FD}_{k,c}\right), \quad \forall k \in \mathcal{K},\label{P4_C1}\\
&\quad\,\,\, \log_2\left(1 + \gamma^{\rm FD}_{k,p}\right) + C_m^{\rm{FD}} \geq \alpha, \quad \forall m \in \mathcal{M}, \quad  \forall k \in \mathcal{K}, \label{P4_C4}\\
&\quad\,\,\, \log_2\left(1 + \gamma^{\rm FD}_{l,p}\right) + C_m^{\rm{FD}} \geq \alpha, \quad \forall m \in \mathcal{M}, \forall l \in \mathcal{L},\label{P4_C4}\\
&\frac{2{\tt{Re}}\left\{\left(\Bar{{\textbf{w}}}_c^{(i)}\right)^H\textbf{h}_{k}^H\textbf{h}_{k}\bar{{\textbf{w}}}_c^{(i)}\right\}}{\mu_{k,c}^{(i)}} - \\
&\frac{\textbf{h}_{k}\bar{{\textbf{w}}}_c^{(i)}\left(\Bar{{\textbf{w}}}_c^{(i)}\right)^H\textbf{h}_{k}^H}{\left(\mu_{k,c}^{(i)}\right)^2} \mu_{k,c},\geq \gamma^{\rm FD}_{k,c},  \forall k \in \mathcal{K}, \label{Aprox13}\\
&\frac{2{\tt{Re}}\left\{\left(\Bar{{\textbf{w}}}_c^{(i)}\right)^H\textbf{h}_{l}^H\textbf{h}_{l}\bar{{\textbf{w}}}_c^{(i)}\right\}}{\mu_{l,c}^{(i)}} - \frac{\textbf{h}_{l}\bar{{\textbf{w}}}_c^{(i)}\left(\Bar{{\textbf{w}}}_c^{(i)}\right)^H\textbf{h}_{l}^H}{\left(\mu_{l,c}^{(i)}\right)^2}\mu_{l,c}+ \notag\\
&\frac{\textbf{h}_{l,d}^H \textbf{v}^{{\rm{FD}},(i)} \left(\textbf{v}^{{\rm{FD}},(i)}\right)^H \textbf{h}_{l,d}}{\sigma^2} + \notag\\
&\frac{2 {\tt{Re}}\left\{\left(\textbf{v}^{{\rm{FD}},(i)}\right)^H \textbf{h}_{l,d} \textbf{h}_{l,d}^H \left( \textbf{v}^{{\rm{FD}}} - \textbf{v}^{{\rm{FD}},(i)} \right)\right\}}{\sigma^2} \geq  \gamma_{d}^{\rm FD}, \forall l \in \mathcal{L},\label{Aprox7}\\
& \frac{2{\tt{Re}}\left\{\left(\Bar{{\textbf{w}}}_{k,p}^{(i)}\right)^H\textbf{h}_{k}^H\textbf{h}_{k}\bar{{\textbf{w}}}_{k,p}^{(i)}\right\}}{\mu_{k,p}^{(i)}} - \notag \\
&\frac{\textbf{h}_{k}\bar{{\textbf{w}}}_{k,p}^{(i)}\left(\Bar{{\textbf{w}}}_{k,p}^{(i)}\right)^H\textbf{h}_{k}^H}{\left(\mu_{k,p}^{(i)}\right)^2}\mu_{k,p} \geq \gamma_{k,p}^{\rm FD} , \forall k \in \mathcal{K},\label{Aprox8}\\
& \frac{2{\tt{Re}}\left\{\left(\Bar{{\textbf{w}}}_{l,p}^{(i)}\right)^H\textbf{h}_{l}^H\textbf{h}_{l}\bar{{\textbf{w}}}_{l,p}^{(i)}\right\}}{\mu_{l,p}^{(i)}} - \notag
\\& \frac{\textbf{h}_{l}\bar{{\textbf{w}}}_{l,p}^{(i)}\left(\Bar{{\textbf{w}}}_{l,p}^{(i)}\right)^H\textbf{h}_{l}^H}{\left(\mu_{l,p}^{(i)}\right)^2}\mu_{l,p} \geq \gamma_{l,p}^{\rm FD} , \forall l \in \mathcal{L},\label{Aprox14}\\
& \mu_{k,c} \ge \textbf{h}_{k}\left(\sum_{m\in\mathcal{M}}\Bar{\textbf{w}}_{k,p}\bar{\textbf{w}}_{k,p}^H\right)\textbf{h}_{k}^H + |h_{k,\rm{SI}}^H v_{k}^{\rm{FD}}|^2 + \sigma^2, \notag \\ & \hspace{5cm} \forall k \in \mathcal{K}, \label{Aprox9}\\
& \mu_{k,p} \ge \textbf{h}_{k}\left(\sum_{j\in\mathcal{M}, j \ne m}\bar{\textbf{w}}_{j,p}\bar{\textbf{w}}_{j,p}^H\right)\textbf{h}_{k}^H + |h_{k,\rm{SI}}^H v_{k}^{\rm{FD}}|^2 + \sigma^2, \notag \\ & \hspace{5cm} \forall k \in \mathcal{K}, \label{Aprox10}\\
& \mu_{l,p} \ge \textbf{h}_{l}\left(\sum_{j\in\mathcal{M}}\bar{\textbf{w}}_{j,p}\bar{\textbf{w}}_{j,p}^H\right)\textbf{h}_{l}^H + \sigma^2, \forall l \in \mathcal{L}, \label{Aprox13}\\
& \mu_{l,c} \ge \textbf{h}_{l}\left(\sum_{m\in\mathcal{M}}\bar{\textbf{w}}_{m,p}\bar{\textbf{w}}_{m,p}^H\right)\textbf{h}_{l}^H + \sigma^2, \forall l \in \mathcal{L},\label{Aprox11}\\
&  \sum_{m \in \mathcal{M}} \left(C_m + \theta g_m\right) - F_{n_b}^{\rm Fh} \le 0,\quad \forall n_b \in \mathcal{N}_b,\\
&   \log_2\left(1+\gamma_{m,p}^{\rm FD, (i)}\right)+\notag\\
&\frac{1}{\left(1+\gamma_{m,p}^{\rm FD, (i)}\right)\ln(2)}\left(\gamma_{m,p}^{\rm FD}-\gamma_{m,p}^{\rm FD, (i)}\right) \le g_m, \forall m \in \mathcal{M}, \label{Aprox12}\\
& \gamma^{\rm FD}_{k,c},\gamma^{\rm FD}_{l,c},   \gamma^{\rm FD}_{k,p}, \gamma^{\rm FD}_{l,p} \geq 0, \forall l \in \mathcal{L},\forall k \in \mathcal{K},
\end{align}
\end{subequations}
\endgroup
where $\boldsymbol{\gamma}^{\rm FD}=[\gamma^{\rm FD}_{k,c}, \gamma^{\rm FD}_{l,c} \gamma^{\rm FD}_{k,p}, \gamma^{\rm FD}_{l,p}, \gamma_{d}^{\rm FD}]$ are introduced to provide a convex representation for the data rate expressions, $\boldsymbol{\mu}=[\mu_{k,p}, \mu_{l,p},\mu_{k,c}, \mu_{l,c}]$ represents the interference plus noise in the SINR calculation. 
We can see that all the constraints in $\tilde{\mathcal{P}}-\mathrm{FD}$ are either linear or convex constraints. Hence, it can be efficiently solved using a standard optimization solver, such as MOSEK.
\bibliographystyle{IEEEtran}

\begin{thebibliography}{10}
\providecommand{\url}[1]{#1}
\csname url@samestyle\endcsname
\providecommand{\newblock}{\relax}
\providecommand{\bibinfo}[2]{#2}
\providecommand{\BIBentrySTDinterwordspacing}{\spaceskip=0pt\relax}
\providecommand{\BIBentryALTinterwordstretchfactor}{4}
\providecommand{\BIBentryALTinterwordspacing}{\spaceskip=\fontdimen2\font plus
\BIBentryALTinterwordstretchfactor\fontdimen3\font minus \fontdimen4\font\relax}
\providecommand{\BIBforeignlanguage}[2]{{%
\expandafter\ifx\csname l@#1\endcsname\relax
\typeout{** WARNING: IEEEtran.bst: No hyphenation pattern has been}%
\typeout{** loaded for the language `#1'. Using the pattern for}%
\typeout{** the default language instead.}%
\else
\language=\csname l@#1\endcsname
\fi
#2}}
\providecommand{\BIBdecl}{\relax}
\BIBdecl

\bibitem{9040264}
M.~Giordani, M.~Polese, M.~Mezzavilla, S.~Rangan, and M.~Zorzi, ``Toward 6g networks: Use cases and technologies,'' \emph{IEEE Communications Magazine}, vol.~58, no.~3, pp. 55--61, 2020.

\bibitem{8869705}
W.~Saad, M.~Bennis, and M.~Chen, ``A vision of 6g wireless systems: Applications, trends, technologies, and open research problems,'' \emph{IEEE Network}, vol.~34, no.~3, pp. 134--142, 2020.

\bibitem{9831440}
Y.~Mao, O.~Dizdar, B.~Clerckx, R.~Schober, P.~Popovski, and H.~V. Poor, ``Rate-splitting multiple access: Fundamentals, survey, and future research trends,'' \emph{IEEE Communications Surveys \& Tutorials}, vol.~24, no.~4, pp. 2073--2126, 2022.

\bibitem{10038476}
B.~Clerckx, Y.~Mao, E.~A. Jorswieck, J.~Yuan, D.~J. Love, E.~Erkip, and D.~Niyato, ``A primer on rate-splitting multiple access: Tutorial, myths, and frequently asked questions,'' \emph{IEEE Journal on Selected Areas in Communications}, vol.~41, no.~5, pp. 1265--1308, 2023.

\bibitem{9123680}
Y.~Mao, B.~Clerckx, J.~Zhang, V.~O.~K. Li, and M.~A. Arafah, ``Max-min fairness of k-user cooperative rate-splitting in miso broadcast channel with user relaying,'' \emph{IEEE Transactions on Wireless Communications}, vol.~19, no.~10, pp. 6362--6376, 2020.

\bibitem{8846761}
J.~Zhang, B.~Clerckx, J.~Ge, and Y.~Mao, ``Cooperative rate splitting for miso broadcast channel with user relaying, and performance benefits over cooperative noma,'' \emph{IEEE Signal Processing Letters}, vol.~26, no.~11, pp. 1678--1682, 2019.

\bibitem{9627180}
T.~Li, H.~Zhang, X.~Zhou, and D.~Yuan, ``Full-duplex cooperative rate-splitting for multigroup multicast with swipt,'' \emph{IEEE Transactions on Wireless Communications}, vol.~21, no.~6, pp. 4379--4393, 2022.

\bibitem{9771468}
S.~Khisa, M.~Almekhlafi, M.~Elhattab, and C.~Assi, ``Full duplex cooperative rate splitting multiple access for a miso broadcast channel with two users,'' \emph{IEEE Communications Letters}, vol.~26, no.~8, pp. 1913--1917, 2022.

\bibitem{9217123}
P.~Li, M.~Chen, Y.~Mao, Z.~Yang, B.~Clerckx, and M.~Shikh-Bahaei, ``Cooperative rate-splitting for secrecy sum-rate enhancement in multi-antenna broadcast channels,'' in \emph{2020 IEEE 31st Annual International Symposium on Personal, Indoor and Mobile Radio Communications}, 2020, pp. 1--6.

\bibitem{7470942}
B.~Clerckx, H.~Joudeh, C.~Hao, M.~Dai, and B.~Rassouli, ``Rate splitting for mimo wireless networks: a promising phy-layer strategy for lte evolution,'' \emph{IEEE Communications Magazine}, vol.~54, no.~5, pp. 98--105, 2016.

\bibitem{9737523}
A.~Mishra, Y.~Mao, L.~Sanguinetti, and B.~Clerckx, ``Rate-splitting assisted massive machine-type communications in cell-free massive mimo,'' \emph{IEEE Communications Letters}, vol.~26, no.~6, pp. 1358--1362, 2022.

\bibitem{8846706}
Y.~Mao, B.~Clerckx, and V.~O.~K. Li, ``Rate-splitting for multi-antenna non-orthogonal unicast and multicast transmission: Spectral and energy efficiency analysis,'' \emph{IEEE Transactions on Communications}, vol.~67, no.~12, pp. 8754--8770, 2019.

\bibitem{9684855}
Z.~W. Si, L.~Yin, and B.~Clerckx, ``Rate-splitting multiple access for multigateway multibeam satellite systems with feeder link interference,'' \emph{IEEE Transactions on Communications}, vol.~70, no.~3, pp. 2147--2162, 2022.

\bibitem{10109654}
S.~Khisa, M.~Elhattab, C.~Assi, and S.~Sharafeddine, ``Energy consumption optimization in ris-assisted cooperative rsma cellular networks,'' \emph{IEEE Transactions on Communications}, vol.~71, no.~7, pp. 4300--4312, 2023.

\bibitem{9676684}
S.~A. Tegos, P.~D. Diamantoulakis, and G.~K. Karagiannidis, ``On the performance of uplink rate-splitting multiple access,'' \emph{IEEE Communications Letters}, vol.~26, no.~3, pp. 523--527, 2022.

\bibitem{9257190}
Z.~Yang, M.~Chen, W.~Saad, W.~Xu, and M.~Shikh-Bahaei, ``Sum-rate maximization of uplink rate splitting multiple access (rsma) communication,'' \emph{IEEE Transactions on Mobile Computing}, vol.~21, no.~7, pp. 2596--2609, 2022.

\bibitem{9445019}
A.~A. Ahmad, Y.~Mao, A.~Sezgin, and B.~Clerckx, ``Rate splitting multiple access in c-ran: A scalable and robust design,'' \emph{IEEE Transactions on Communications}, vol.~69, no.~9, pp. 5727--5743, 2021.

\bibitem{8756076}
D.~Yu, J.~Kim, and S.-H. Park, ``An efficient rate-splitting multiple access scheme for the downlink of c-ran systems,'' \emph{IEEE Wireless Communications Letters}, vol.~8, no.~6, pp. 1555--1558, 2019.

\bibitem{9573421}
J.~Zhou, Y.~Sun, R.~Chen, and C.~Tellambura, ``Rate splitting multiple access for multigroup multicast beamforming in cache-enabled c-ran,'' \emph{IEEE Transactions on Vehicular Technology}, vol.~70, no.~12, pp. 12\,758--12\,770, 2021.

\bibitem{9759225}
K.~Weinberger, A.~A. Ahmad, A.~Sezgin, and A.~Zappone, ``Synergistic benefits in irs- and rs-enabled c-ran with energy-efficient clustering,'' \emph{IEEE Transactions on Wireless Communications}, vol.~21, no.~10, pp. 8459--8475, 2022.

\bibitem{9896157}
Q.~Zhu, Z.~Qian, X.~Gao, and X.~Wang, ``Analytical modeling of rsma-enabled user-centric rrh clustering in c-ran over generalized fading channels,'' \emph{IEEE Wireless Communications Letters}, vol.~11, no.~12, pp. 2550--2554, 2022.

\bibitem{8756668}
Y.~Mao, B.~Clerckx, and V.~O.~K. Li, ``Rate-splitting multiple access for coordinated multi-point joint transmission,'' in \emph{2019 IEEE International Conference on Communications Workshops (ICC Workshops)}, 2019, pp. 1--6.

\bibitem{9195473}
J.~Zhang, J.~Zhang, Y.~Zhou, H.~Ji, J.~Sun, and N.~Al-Dhahir, ``Energy and spectral efficiency tradeoff via rate splitting and common beamforming coordination in multicell networks,'' \emph{IEEE Transactions on Communications}, vol.~68, no.~12, pp. 7719--7731, 2020.

\bibitem{9461768}
Z.~Yang, M.~Chen, W.~Saad, and M.~Shikh-Bahaei, ``Optimization of rate allocation and power control for rate splitting multiple access (rsma),'' \emph{IEEE Transactions on Communications}, vol.~69, no.~9, pp. 5988--6002, 2021.

\bibitem{9508885}
A.~Bansal, K.~Singh, B.~Clerckx, C.-P. Li, and M.-S. Alouini, ``Rate-splitting multiple access for intelligent reflecting surface aided multi-user communications,'' \emph{IEEE Transactions on Vehicular Technology}, vol.~70, no.~9, pp. 9217--9229, 2021.

\bibitem{9854887}
S.~Dhok and P.~K. Sharma, ``Rate-splitting multiple access with star ris over spatially-correlated channels,'' \emph{IEEE Transactions on Communications}, vol.~70, no.~10, pp. 6410--6424, 2022.

\bibitem{10436906}
S.~Khisa, M.~Elhattab, C.~Assi, and S.~Sharafeddine, ``Ris-assisted swipt-empowered cooperative rate-splitting multiple access for two users,'' in \emph{GLOBECOM 2023 - 2023 IEEE Global Communications Conference}, 2023, pp. 7231--7236.

\bibitem{10233866}
C.~Kong and H.~Lu, ``Cooperative rate-splitting multiple access in heterogeneous networks,'' \emph{IEEE Communications Letters}, vol.~27, no.~10, pp. 2807--2811, 2023.

\bibitem{mao2018rate}
Y.~Mao, B.~Clerckx, and V.~O. Li, ``Rate-splitting multiple access for downlink communication systems: Bridging, generalizing, and outperforming sdma and noma,'' \emph{EURASIP journal on wireless communications and networking}, vol. 2018, pp. 1--54, 2018.

\bibitem{8352643}
M.~S. Ali, E.~Hossain, A.~Al-Dweik, and D.~I. Kim, ``Downlink power allocation for comp-noma in multi-cell networks,'' \emph{IEEE Transactions on Communications}, vol.~66, no.~9, pp. 3982--3998, 2018.

\bibitem{9382277}
Y.~Mao, E.~Piovano, and B.~Clerckx, ``Rate-splitting multiple access for overloaded cellular internet of things,'' \emph{IEEE Transactions on Communications}, vol.~69, no.~7, pp. 4504--4519, 2021.

\bibitem{10154612}
Y.~Sun, Z.~Ding, X.~Dai, M.~Zhou, and Z.~Ding, ``On the application of quasi-degradation to network noma in downlink comp systems,'' \emph{IEEE Transactions on Wireless Communications}, vol.~23, no.~2, pp. 978--993, 2024.

\bibitem{7839266}
S.~Bassoy, H.~Farooq, M.~A. Imran, and A.~Imran, ``Coordinated multi-point clustering schemes: A survey,'' \emph{IEEE Communications Surveys \& Tutorials}, vol.~19, no.~2, pp. 743--764, 2017.

\bibitem{8302918}
X.~Yue, Y.~Liu, S.~Kang, A.~Nallanathan, and Z.~Ding, ``Spatially random relay selection for full/half-duplex cooperative noma networks,'' \emph{IEEE Transactions on Communications}, vol.~66, no.~8, pp. 3294--3308, 2018.

\bibitem{8094966}
G.~Liu, X.~Chen, Z.~Ding, Z.~Ma, and F.~R. Yu, ``Hybrid half-duplex/full-duplex cooperative non-orthogonal multiple access with transmit power adaptation,'' \emph{IEEE Transactions on Wireless Communications}, vol.~17, no.~1, pp. 506--519, 2018.

\bibitem{access2012radio}
E.~U. T.~R. Access, ``Radio resource control (rrc), 3gpp ts 36.331,'' 2012.

\bibitem{6195835}
W.~Cheng, X.~Zhang, and H.~Zhang, ``Full/half duplex based resource allocations for statistical quality of service provisioning in wireless relay networks,'' in \emph{2012 Proceedings IEEE INFOCOM}, 2012, pp. 864--872.

\bibitem{8491100}
Y.~Mao, B.~Clerckx, and V.~O. Li, ``Energy efficiency of rate-splitting multiple access, and performance benefits over sdma and noma,'' in \emph{2018 15th International Symposium on Wireless Communication Systems (ISWCS)}, 2018, pp. 1--5.

\bibitem{8648507}
H.~M. Al-Obiedollah, K.~Cumanan, J.~Thiyagalingam, A.~G. Burr, Z.~Ding, and O.~A. Dobre, ``Energy efficient beamforming design for miso non-orthogonal multiple access systems,'' \emph{IEEE Transactions on Communications}, vol.~67, no.~6, pp. 4117--4131, 2019.

\bibitem{beck2010sequential}
A.~Beck, A.~Ben-Tal, and L.~Tetruashvili, ``A sequential parametric convex approximation method with applications to nonconvex truss topology design problems,'' \emph{Journal of Global Optimization}, vol.~47, pp. 29--51, 2010.

\bibitem{9737471}
M.~Elhattab, M.~A. Arfaoui, and C.~Assi, ``Joint clustering and power allocation in coordinated multipoint assisted c-noma cellular networks,'' \emph{IEEE Transactions on Communications}, vol.~70, no.~5, pp. 3483--3498, 2022.

\end{thebibliography}

\end{document}